\documentclass[%
 reprint,
superscriptaddress,
amsmath,amssymb,
aps,
prx,
]{revtex4-2}

\usepackage{graphicx}
\usepackage{dcolumn}
\usepackage{bm}
\usepackage[colorlinks]{hyperref}
\usepackage{physics}
\usepackage[squaren]{SIunits}
\usepackage{mathtools}
\usepackage{amsthm}
\usepackage{xcolor}
\usepackage{mathrsfs}
\usepackage{times}
\usepackage[normalem]{ulem}
\usepackage{algorithm2e}
\usepackage{comment}
\usepackage{titletoc}

\usepackage{tocloft}

\hypersetup{
	colorlinks=true,
	linkcolor=red,
	filecolor=blue,      
	urlcolor=blue,
	citecolor=blue
}
\newcommand{\rev}[1]{{\textcolor{black}{#1}}}

\newcommand{\ci}{\mathbf{V}}       
\newcommand{\gr}{\mathbf{G}}       

\begin{document}

\title[]{Generalization Error in Quantum Machine Learning in the Presence of Sampling Noise}


\makeatletter

\author{Fangjun Hu}
\email{fhu@princeton.edu}
\affiliation{Department of Electrical and Computer Engineering, Princeton University, Princeton, NJ 08544, USA}

\author{Xun Gao}
\affiliation{JILA and Department of Physics, University of Colorado Boulder, Boulder, CO 80309, USA}

\date{\today}

\begin{abstract}
    Tackling output sampling noise due to finite shots of quantum measurement is an unavoidable challenge when extracting information in machine learning with physical systems. \rev{A technique called} Eigentask Learning was developed recently as a framework for \rev{learning with infinite input training data} in the presence of output sampling noise. In the work of Eigentask Learning, numerical evidence was presented that extracting low-noise contributions of features can \rev{practically} improve performance for machine learning tasks, displaying robustness to overfitting and increasing generalization accuracy. \rev{However, it remains unsolved to quantitatively characterize generalization errors in situations where the training dataset is finite, while output sampling noise still exists.} In this study, we use methodologies from statistical mechanics to calculate the training and generalization errors of a generic quantum machine learning system when the input training dataset and output measurement sampling shots are both finite. Our analytical findings, supported by numerical validation, offer solid justification that Eigentask Learning provides optimal learning in the sense of minimizing generalization errors.
\end{abstract}

\maketitle

\section{Introduction}

Exploring quantum machine learning (QML) schemes that leverage underlying quantum systems for processing classical data has been studied in various applications \cite{grant_hierarchical_2018, tacchino_quantum_2020, chen_temporal_2020, hu_overcoming_2024, cong_quantum_2019}.
One particular way of applying quantum systems to supervised machine learning algorithms is called parametric quantum circuits (PQC) \cite{sim_expressibility_2019, larose_robust_2020, schuld_effect_2021, Abbas_power_2021, wu_expressivity_2021, wright_capacity_2019, Du2020, Holmes2022, farhi_quantum_2014, farhi_classification_2018, schuld_circuit_2020, cong_quantum_2019}, where the quantum circuits depend not only on the task inputs but also some internal parameters.
\rev{Making predictions using PQCs requires accessing information from these circuits. To achieve this, one can measure all qubits in the PQC under computational bases and get the probabilities of obtaining classical bitstrings.
These probabilities, as functions of inputs, are called \textit{features} in QML. Then, we can use the linear combination of these features through a set of \textit{output weights} to make predictions for any given input. In generic PQC schemes, both the internal parameters and output weights need to be trained.}

\rev{However, optimizing the internal parameters inside PQC is usually difficult and costly.} It is a challenge being widely investigated \cite{mcclean_barren_2018, cerezo_cost_2021, cerezo_provable_2023}. To circumvent this difficulty, QML schemes without optimizing PQC internal parameters, including quantum kernel methods \cite{havlicek_supervised_2019}, quantum extreme learning machine \cite{innocenti_potential_2022}, and quantum reservoir computing (QRC) \cite{suzuki_natural_2022, hu_overcoming_2024}, are proposed. 
Among these schemes, QRC is a QML scheme that maps the classical input data to a high-dimensional vector in the space of features, and only the output weights are optimized, leaving all randomly initialized internal parameters untrained. 
\rev{It can be implemented in NISQ devices on many tasks more easily  \cite{hu_overcoming_2024, kornjaca_large_2024}. Therefore, we will restrict our discussion to QRC schemes in this article. }

Even in an ideal quantum computer, we emphasize that QML contains two main types of stochasticities. \rev{The first one is the randomness of input samples appearing in any machine learning algorithms with both classical and quantum systems.} Specifically, in practice, the learning process in any supervised QML algorithms must be carried out on a finite training input dataset, which can be considered drawn from some prior distribution. 
The evaluation of the trained weights and parameters based on this finite training input dataset is directly related to two important quantities in classical machine learning: training and generalization errors \cite{canatar_spectral_2021}.
The second stochasticity is called quantum sampling noise (QSN) \cite{hu_tackling_2023} -- the statistical fluctuations during repeated measurement \textit{shots} of quantum systems -- on the output side. 
In contrast to classical machine learning, QML suffers from this inherent QSN which places a strong constraint on practical optimization and performance \cite{arrasmith_effect_2021, garcia-beni_scalable_2022}. 

A recently developed theoretical framework \cite{hu_tackling_2023}, Eigentask Analysis, presents a canonical representation of the quantum reservoir computing schemes and rigorously quantifies the impact of QSN. 
\rev{The goal of Eigentask Analysis is to find functions restricted as specific linear combinations of original features so that these eigentasks are minimally affected by QSN. Those newly constructed functions are referred as \textit{Eigentasks} in Ref.\,\cite{hu_tackling_2023}. 
Here, restricting eigentasks to be the linear combination of original features does not change the essence of quantum reservoir computing. This is because it is equivalent to training output weights either using the combination of all original features or using the combination of all eigentasks.}
Thereafter, Ref.\,\cite{hu_tackling_2023} provided numerical evidence that maximal testing accuracy can be achieved by truncating specific high-noise eigentasks. This technique was referred as Eigentask Learning. The motivation of Eigentask Learning is that eigentasks with high noise are irrelevant or even harmful for learning. \rev{However, a more comprehensive theoretical understanding of such a phenomenon was unclear before \cite{hu_tackling_2023, Polloreno2022}, and we now solve this problem in this paper.}

Various mathematical techniques derived from statistical mechanics have found applications in computing generalization error in classical machine learning \cite{Sompolinsky1990, seung_statistical_1992, canatar_spectral_2021, canatar_statistical_2022, Advani2013, Crisanti2018}. This connection stems from the observation that the equilibrium distribution of a stochastic gradient descent training process corresponds to the Gibbs distribution in thermodynamics \cite{seung_statistical_1992}. And the corresponding free energy plays a role of generalization error after training.
Therefore, these previous approaches analogize trained input data and training weights as quenched disorder (the coupling strength) and spin configuration, respectively, in a spin-glass system. 
This analogy enables a quantitative analysis of the performance of QML systems when encountering new, unseen data after being trained on finite datasets. 
\rev{In this article, we generalize this methodology to QRC, where both input data and output samples are finite. These two randomnesses constitute the newly introduced analog of quenched disorder.} Specifically, this paper formally defines training and generalization errors as functions of input dataset size $N$ and the shot number $S$ in QRC. We then solve them rigorously by using replica methods in spin-glass theory. 
Based on these results, we formally justify that a well-defined truncation of eigentasks minimizes the generalization error in QRC, the phenomenon previously lacking a theoretical explanation in Ref.\,\cite{hu_tackling_2023}. To the best of our knowledge, this work is the first to analyze training and generalization while accounting for both input and output sampling noise in QML.

This article is organized as follows. Sec.\,\ref{sec:Theoretical_Setup} provides a theoretical setup of utilizing quantum systems for learning, with a brief review of the concepts of eigentask. Sec.\,\ref{sec:Training_Generalization} formally introduce the definitions of training and generalization errors. Those two errors evaluated from the numerical simulation are compared to the theoretical result derived by the replica trick. The close alignment observed in this comparison shows the validity of the analytical methods. Sec.\,\ref{sec:Eigentask_Learning} applies the results established in Sec.\,\ref{sec:Training_Generalization} to prove the optimal order-truncation of eigentasks, laying a rigorous foundation of Eigentask Learning. 

\section{Eigentask Analysis for Learning with Quantum Systems}
\label{sec:Theoretical_Setup}

The underlying part of any QML architecture is a quantum system with an associated Hilbert space dimension $K$. For an $L$-qubit quantum system, the dimension of the quantum system is $K = 2^L$. This quantum system encodes classical input data into the quantum system and outputs features through measurement. 

To be more specific, we first let the \textit{input dataset} $\{\bm{u}^{(n)} \in \mathbb{R}^D \}_{n \in [N]}$ be $N$ i.i.d.~$D$-dimensional classical random variables which are drawn from a certain distribution $p (\bm{u})$. It naturally induces an expectation over the input $\mathbb{E}_{\bm{u}} [f] = \int f (\bm{u}) p (\bm{u}) \dd \bm{u}$ for any function $f$. Being initialized to $\hat{\rho}_0$, the quantum system encodes any input $\bm{u}$ into a quantum circuit $\hat{U} (\bm{u})$, and then the initial state evolves to a new quantum state $\hat{\rho} (\bm{u}) = \hat{U} (\bm{u}) \hat{\rho}_0 \hat{U} (\bm{u})^\dagger$. To make information accessible, we assign a set of positive operator-valued measure (POVM) elements $\{ \hat{M}_k \}_{k \in [K]}$. One commonly chosen measurement in qubit system is through the projective basis $\hat{M}_k = \ket{k} \bra{k},$ \rev{where $k$ is in binary representation.} The direct readout feature functions are the quantum probabilities of obtaining bitstring $k$, conditioned on the input $\bm{u}$: $x_k (\bm{u}) = \mathrm{Tr} (\hat{M}_k \hat{\rho} (\bm{u})) = \mathrm{Pr} [k | \bm{u} ]$. We denote the collection of all features $\bm{x}(\bm{u}) = \{x_k(\bm{u})\}_{k \in [K]}$ which also has a dimension $K$. 

In practice, the conditional probability above is usually inaccessible due to the finite number of quantum measurement repetitions or shots, denoted as $S$. For the input $\bm{u}$, in the $s$-th shot the quantum measurement only produces one outcome $k^{(s)} (\bm{u})$. 
\rev{It is convenient to define \textit{output sample sequence} (of a single $\bm{u}$) $\mathcal{X} (\bm{u}) = \{ k^{(s)} (\bm{u}) \}_{s \in [S]}$ and the \textit{overall dataset}}
\begin{equation}
    \mathcal{D}= \{ (\bm{u}^{(n)}, \mathcal{X} (\bm{u}^{(n)})) \}_{n \in [N]} .
\end{equation}
From the perspective of statistical mechanics, similar to the case of spin-glass, we will see in Appendix \ref{apx:Replica_trick} the overall dataset $\mathcal{D}$ also serves as a quenched disorder in the (quantum) machine learning system.
The emprical \textit{noisy readout features} $\bm{X}(\bm{u}) = \{ X_k(\bm{u}) \}_{k \in [K]}$ counts the occurrence of each $k$
\begin{equation}
    X_k (\bm{u}) = \frac{1}{S} \sum_{s = 1}^S \delta (k^{(s)} (\bm{u}), k) 
    = x_k (\bm{u}) + \frac{1}{\sqrt{S}} \zeta_k (\bm{u}), \label{eq:qsn}
\end{equation}
where $\delta$ is the indicator function, and $\bm{\zeta}$ is a zero-mean random variable, with covariance $\mathbf{\Sigma} (\bm{u}) := \mathrm{Cov}_{\mathcal{X} (\bm{u})} [\bm{\zeta}] = \mathrm{diag} (\bm{x}) - \bm{x} \bm{x}^T$. Here we use $\mathbb{E}_{\mathcal{X} (\bm{u})}$ and $\mathrm{Cov}_{\mathcal{X} (\bm{u})}$ as the conditional expectation and covariance over output samples $\mathcal{X} (\bm{u})$, conditioned on the input value $\bm{u}$. \rev{The term $\frac{1}{\sqrt{S}} \zeta_k (\bm{u})$ in Eq.\,(\ref{eq:qsn}) is called \textit{quantum sampling noise} (QSN).}

Theory of Eigentask Analysis \cite{hu_tackling_2023} shows that there exists a unique $S$-independent tuple $\{ \beta_k^2, \bm{r}^{(k)} \}$ which simultaneously diagonalizes both new signals $y^{(k)} = \sum_j r_j^{(k)} x_j$ and new noises $\xi^{(k)} = \sum_j r_j^{(k)} \zeta_j$, such that 
\begin{equation}
    \left\{\begin{array}{l}
        \mathbb{E}_{\boldsymbol{u}} [y^{(k)} y^{(k')}] = \delta_{k k'}, \\
        \mathbb{E}_{\boldsymbol{u}} [\mathbb{E}_{\mathcal{X}(\bm{u})} [\xi^{(k)} \xi^{(k')}] ] = \beta_k^2 \delta_{k k'}.
    \end{array}\right.
\end{equation}
Here, the tuple $\{ \beta_k^2, \bm{r}^{(k)} \}$ can be uniquely solved via the generalized eigenproblem
\begin{align}
    \ci \bm{r}^{(k)} = \beta^2_k \gr \bm{r}^{(k)}, \label{eq:VrGr}
\end{align}
where $\gr =\mathbb{E}_{\bm{u}} [\bm{x} \bm{x}^T]$ is the \textit{Gram matrix} and $\ci = \mathbb{E}_{\bm{u}} [\mathbf{\Sigma}]$ is the \textit{mean noise matrix}. The newly constructed signals $\{ y^{(k)} \}$ are called \textit{eigentasks}, and $\{ \beta^2_k \}$ are their associated eigen-noise-to-signal-ratios (eigen-NSRs). We use a non-decreasing order $0 = \beta^2_1 \leq \beta^2_2 \leq \cdots \leq \beta^2_K < \infty$ to label those eigentasks and eigen-NSRs. We will show in the next section that the training and generalization errors can be fully captured merely in terms of $N,S$ and $\{ \beta_k^2 \}$, under the eigentask representation.
\rev{We also emphasize that, in Ref.\,\cite{hu_tackling_2023}, it is shown that even if these $\{\beta_k^2, \bm{r}^{(k)}\}$ are defined through noiseless features $\bm{x}(\bm{u})$, they can be accurately estimated from noisy features $\bm{X}(\bm{u})$ through a certain empirical correction (see Appendix D in Ref.\,\cite{hu_tackling_2023}).}

\section{Training and Generalization Error}
\label{sec:Training_Generalization}

The basic task of quantum reservoir computing is using the linear combination of features $\bm{X}(\bm{u})$ to fit some given (non-random) target function $f^{\star} (\bm{u})$. We let $f^{\star}$ have a decomposition 
\begin{align}
    f^{\star} = \bm{c} \cdot \bm{x} + f_{\bot}, \label{eq:cx}
\end{align}
where $\int f_{\bot} (\bm{u}) x_k (\bm{u}) \dd \bm{u} = 0$ for all $k$ and this $f_{\bot}$ captures the upper limit of the prediction performance of the quantum system. 
\rev{It is also convenient to decompose $f^{\star}$ under the basis of eigentasks
\begin{align}
    f^{\star} = \bm{a} \cdot \bm{y} + f_{\bot}. \label{eq:ay}
\end{align}
The transform between $\bm{a}$ and $\bm{c}$ are through the $\{\bm{r}^{(k)}\}$ coefficients computed in Eq.\,(\ref{eq:VrGr}). Such transform exists in any generic quantum reservoir without particular symmetry due to the linear independence of $\{\bm{r}^{(k)}\}$ (see Appendix C5 in Ref.\,\cite{hu_tackling_2023}). }

Without loss of generality, we assume $f^{\star}$ is normalized $\mathbb{E}_{\bm{u}} [f^{\star 2}] = \bm{c}^{T} \mathbf{G} \bm{c} +\mathbb{E}_{\bm{u}} [f^2_{\bot}] = 1$. Conditioned on overall dataset (or quenched disorder) $\mathcal{D}$, the training process is to minimize the loss function $H (\bm{w}) = \frac{1}{2 \lambda} \sum_{n} \mathcal{L} (\bm{w}, \mathcal{X} (\bm{u}^{(n)})) + \frac{1}{2} \| \bm{w} \|^2$ over all possible \textit{output weights} $\bm{w} \in \mathbb{R}^K$, where single-input loss function is $\mathcal{L} (\bm{w}, \mathcal{X} (\bm{u})) = (\bm{w} \cdot \bm{X} (\bm{u}) - f^{\star} (\bm{u}))^2$. We emphasize that $\mathcal{L}$ is also stochastic due to the finiteness of $S$. The term $\frac{\lambda}{2} \| \bm{w} \|^2$ with small parameter $\lambda$ is added as a regularizer to ensure the uniqueness of optimized $\bm{w}^{\ast}$ \cite{canatar_spectral_2021}. The training procedure finally yields unique $\mathcal{D}$-dependent \textit{trained weights} 
\begin{align}
    \bm{w}^{\ast} = \underset{\bm{w} \in \mathbb{R}^K}{\mathrm{argmin}} ~ H (\bm{w}). 
\end{align} 
Ideally, if $N,S$ tends to infinite, $\bm{w}^{\ast}$ tends to $\bm{c}$ for small enough regularizer $\lambda$. 
The training error function is directly defined as the loss by setting output weights be the trained weights:
\begin{equation}
    \epsilon_t (\bm{w}^{\ast} ) = \frac{1}{N}  \sum_{n = 1}^N \mathcal{L} (\bm{w}^{\ast} , \mathcal{X} (\bm{u}^{(n)})). \label{eq:def_epsilon_t}
\end{equation}
The definition of the generalization error function of weights $\bm{w}$ is a bit more involved. We define it as $\epsilon_g (\bm{w}) = \mathbb{E}_{\bm{u}} [\mathbb{E}_{\mathcal{X} (\bm{u})} [\mathcal{L} (\bm{w}, \mathcal{X} (\bm{u}))]]$. A detailed calculation (see Appendix \ref{apx:Generalization_Function}) gives the explicit form of $\epsilon_g$
\begin{equation}
    \epsilon_g (\bm{w}) = \bm{w}^T \left( \mathbf{G}+ \frac{1}{S} \mathbf{V} \right) \bm{w} - 2 \bm{c}^{T} \mathbf{G} \bm{w} + 1. \label{eq:def_epsilon_g}
\end{equation}
Generalization error $\epsilon_g$ evaluates how well the quantum learning system performs for unseen data using the output weights $\bm{w}$. It can be justified in the following sense. Suppose that quantum system sees a new testing dataset $\mathcal{D}' = \{ (\bm{v}^{(n')}, \mathcal{X} (\bm{v}^{(n')})) \}_{n' \in [N']}$, which is independent from $\mathcal{D}$, but still obeys $\bm{v}^{(n')} \sim p(\bm{v})$. The testing error $\epsilon_{\mathrm{ts}} (\bm{w}) = \frac{1}{N'} \sum_{n'} (\bm{w} \cdot \bm{X} (\bm{v}^{(n')}) - f^{\star} (\bm{v}^{(n')}))^2$ can be shown to converge to the deterministic quantity in Eq.\,(\ref{eq:def_epsilon_g}) when $N' \to \infty$ (see Appendix \ref{apx:Generalization_Function}).

However, quantities $\epsilon_t(\bm{w}^\ast)$ and $\epsilon_g(\bm{w}^\ast)$ are both $\mathcal{D}$-dependent and therefore random. To unambiguously quantify the training and generalization errors, we need to take the ensemble average over all possible $\mathcal{D}$. It leads to the central concepts of this work, the \textit{average} training and generalization errors, namely
\begin{equation}
    E_t =\mathbb{E}_{\mathcal{D}} [\epsilon_t (\bm{w}^{\ast} )], \quad
E_g =\mathbb{E}_{\mathcal{D}} [\epsilon_g (\bm{w}^{\ast})].
\end{equation}

\begin{figure}
    \centering
    \includegraphics[width=\linewidth]{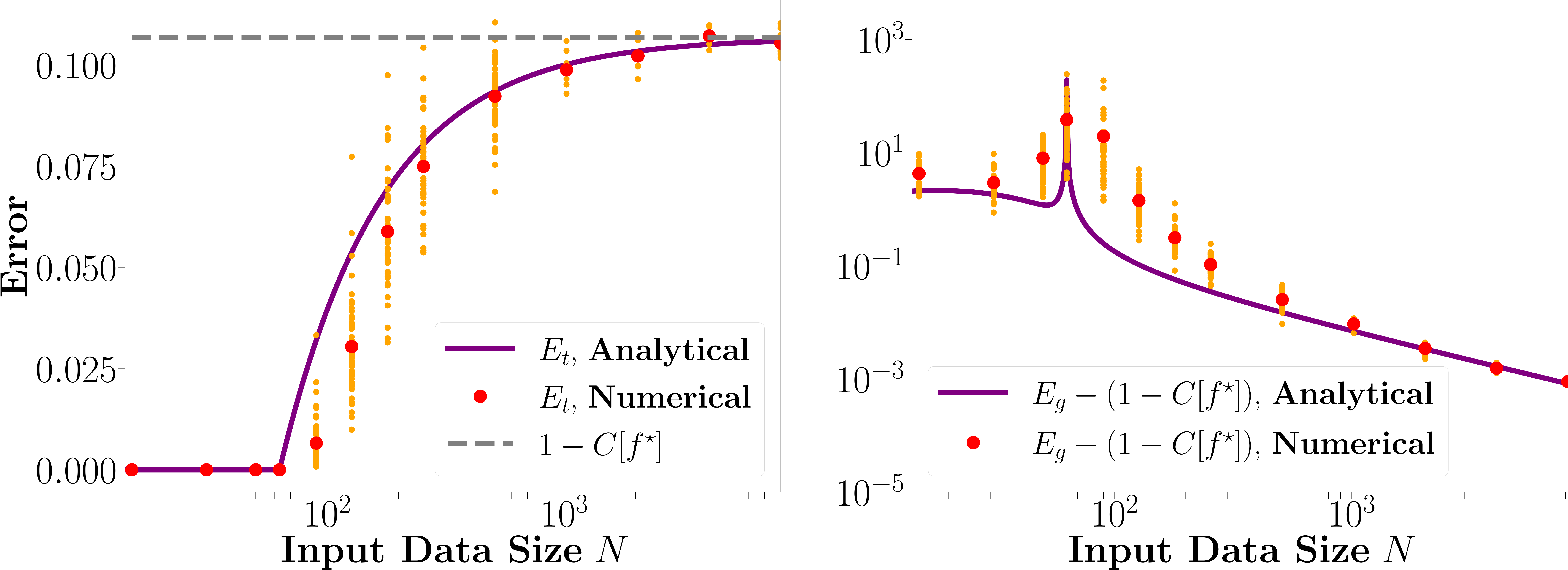}
    \caption{Learning curves for fitting a 1D function in a $6$-qubit system. The target function is $f^\star(u) = \mathrm{Sgn}(u)$ where $u \in [-1, 1]$. The remaining parameters are $K - 1 = 63, S = 10^3, \lambda=10^{-5}$ and $\tau = 2$. (Left) Training error $E_t$ as a function of training dataset size $N$. (Right) Generalization error excess $E_g - (1- C[f^\star])$ as a function of $N$; the y-axis of this plot is log-scaled. For each value of $N$, $R$ different independent overall datasets $\mathcal{D}$ are generated, where $R=50$ for $N \leq 1000$ and $R=10$ for $N > 1000$. The empirical training and generalization errors of each dataset are dots in orange. The mean values of empirical errors (red dots), and the theoretical error values computed from Eq.\,(\ref{eq:Et}) and Eq.\,(\ref{eq:Eg}) (purple line) are compared at each $N$. The dashed gray line in the left panel is $1-C[f^\star]$, the large $N$ limit for both training and generalization errors. }
    \label{fig:1}
\end{figure}

By using the replica trick, we can show that (see Appendix \ref{apx:Original_Features} and \ref{apx:Eigentask}) the average training and generalization errors \rev{under eigentask representation} are
\begin{align}
    E_t & = \frac{\lambda^2}{\kappa^2} E_g, \label{eq:Et} \\
    E_g & = \frac{1}{1 - \gamma} \!\! \left( \! \mathbb{E}_{\bm{u}} [f^2_{\bot}] + \! \sum_k a_k^{2} \!\! \left( \! \frac{\left( \frac{\beta_k^2}{S} + \frac{\kappa}{N} \right)^2 + \frac{\beta_k^2}{S}}{\left( 1 + \frac{\beta_k^2}{S} + \frac{\kappa}{N} \right)^2} \! \right) \!\! \right)\!\!,  \label{eq:Eg}
\end{align}
where there are two new quantities we need to specify:
\begin{align}
    \kappa & = \lambda + \kappa \sum_{k > 1} \frac{1 + \beta_k^2/S}{N ( 1 + \beta_k^2/S ) + \kappa}, \label{eq:kappa}\\
    \gamma & = \sum_{k > 1} \frac{N ( 1 + \beta_k^2/S )^2}{( N ( 1 + \beta_k^2/S ) + \kappa )^2}.
\end{align}
The value of $\kappa$ must be solved self-consistently from Eq.\,(\ref{eq:kappa}). It is usually called Signal Capture Threshold (SCT) \cite{jacot_machine_2020} in classical machine learning theory.

There are three comments that can justify the correctness of the formula Eq.\,(\ref{eq:Et}-\ref{eq:Eg}). First of all, by taking the noiseless limit $S \to \infty$, we get $E_g = \frac{1}{1 - \gamma} ( \mathbb{E}_{\bm{u}} [f^2_{\bot}] + \sum_{k > 1} a_k^{2} ( \frac{\kappa}{\kappa + N} )^2 )$, which reproduces results of generalization error in classical kernel regression with unit and flat Reproducing Kernel Hilbert Space (RKHS) spectrum \cite{canatar_spectral_2021, canatar_statistical_2022}. Secondly, when the regularization parameter $\lambda$ is small enough, in the limit of large training dataset $N \to \infty$, the quantum system sees enough data so that it learns sufficiently, and there should be no difference between training and generalization errors:
\begin{equation}
    \lim_{N \to \infty} E_t (\lambda = 0) = \lim_{N \to \infty} E_g (\lambda = 0) = 1 - C [f^{\star}],
\end{equation}
where $C [f^{\star}] = \sum_k \frac{a_k^{2}}{1 + \beta_k^2 / S}$ is called the \textit{functional capacity} of $f^\star$, defined in Ref.\,\cite{hu_tackling_2023}. Finally, also for $\lambda \to 0$ limit, we can show that $\kappa$ is zero if $N$ is larger than $K - 1$, while $\kappa$ is non-zero if $N$ is smaller than $K - 1$ (see Appendix \ref{apx:Eigentask}).
It hints a phase transition at $N=K-1$. 
\rev{When $N < K-1$, the training error $E_t$ is shown to be zero for $\lambda \to 0$. This is because the quantum systems do not see enough data and overfit the data point by interpolation. As a result, the generalization error $E_g$ maintains a relatively high value and reaches its maximum at $N = K-1$. On the other hand, when $N > K-1$, the quantum system can no longer interpolate; therefore, $E_t$ starts increasing while $E_g$ decreases.}
\rev{In practice, $K=2^L$ is exponentially large and is typically larger than $N$. As a result, if one does not properly drop some features, then this quantum reservoir computing algorithm is always in the interpolation phase $N<K-1$. This observation means a certain dropping of features is crucial for learning, which will be discussed in Sec.\,\ref{sec:Eigentask_Learning}.}

Fig.\,\ref{fig:1} shows the training error $E_t$, and the logarithm of deviation between generalization error $E_g$ and its large $N$ limit $1-C[f^\star]$, when fitting target function $f^{\star} (u) = \mathrm{Sgn} (u)$, where 1D input $u \in [- 1, 1]$. Here, $\mathrm{Sgn} (u) = -1$ for $u<0$ and $\mathrm{Sgn} (u) = 1$ for $u\geq 0$. The underlying quantum system is a $6$-qubit quantum Ising model with a random drive. More concretely, the Hamiltonian is $\hat{H} (u) = \hat{H}_0 + u \cdot \hat{H}_1$, where $\hat{H}_0 = \sum_{\langle l, l' \rangle} J \hat{\sigma}_l^z \hat{\sigma}_{l'}^z + \sum_l (h_l^x \hat{\sigma}_l^x + h_l^z \hat{\sigma}_l^z)$ and $\hat{H}_1 = \sum_l h_l^I \hat{\sigma}_l^x$, where all $h_l^{x/z/I}$ are randomly drawn through uniform distribution of certain intervals. The encoding is $\hat{\rho} (u) = e^{- i \tau \hat{H} (u)} \hat{\rho}_0 e^{i \tau \hat{H} (u)}$. Both empirical average training and generalization errors and their theoretical error values computed from Eq.\,(\ref{eq:Et}) and Eq.\,(\ref{eq:Eg}) are shown in Fig.\,\ref{fig:1}. These data match well, especially for the practical large $N$ limit. This is because in the calculation of replica methods, one needs to take the saddle point approximation and assume $N$ to be large (see Appendix \ref{apx:rsa} and \ref{apx:spe}). Therefore, the larger $N$ is, the more accurate the replica theory produces. 

We find that our numerical is consistent with the theoretical prediction of phase transition at $N=K-1$. 
When $N<K-1$, the quantum reservoir interpolates so that $E_t$ is nearly zero but $E_g$ is high. 
Then, when $N > K-1$, training errors start to become non-zero since perfect interpolation is no longer possible for $N > K-1$. At the same time, generalization errors $E_g$ also decrease until both $E_t$ and $E_g$ converge to their common limit $1-C[f^\star]$. Such phenomenon in generalization errors is called \textit{double-descent}, it was also observed in many classical machine learning systems \cite{canatar_spectral_2021, Li2020}.

\section{Eigentask Learning}
\label{sec:Eigentask_Learning}

\rev{As we discussed in Sec.\,\ref{sec:Training_Generalization}, the feature number $K$ in QML is usually larger than the input training dataset size $N$, which corresponds to the interpolation phase. Therefore, truncating the number of used features for training and testing is necessary. Ref.\,\cite{hu_tackling_2023} proposed a learning scheme called Eigentask Learning. It keeps the leading $K_L$ eigentasks $\{ y^{(k)} \}_{k \in [K_L]}$ with lowest-noise to do the training and testing. This section will follow this methodology and determine its optimal truncation.}

First of all, we take the limitation $\lambda \rightarrow 0$ in Eq.\,(\ref{eq:Eg}) and fix the training dataset size $N$. \rev{Then for any $K_L - 1 < N$, by solving self-consistent equation of $\kappa$, we have $\kappa = 0$ and $\gamma = (K_L - 1)/N$ (see Appendix \ref{apx:Eigentask}). Then, the average generalization error in this case is
\begin{align}
    E_g =~& \frac{N}{N - K_L - 1} \Bigg(\sum_{k=1}^{K_L} a_k^{2} \left( \frac{\beta_k^2 / S}{1 + \beta_k^2 / S} \right) \nonumber\\
    & + \sum_{k=K_L+1}^{K} a_k^{2} + \mathbb{E}_{\bm{u}} [f^2_{\bot}] \Bigg). \label{eq:EgKL}
\end{align}
Now we can explain the efficacy of Eigentask Learning, by using Eq.\,(\ref{eq:EgKL}).}

\begin{figure}
    \centering
    \includegraphics[width=\linewidth]{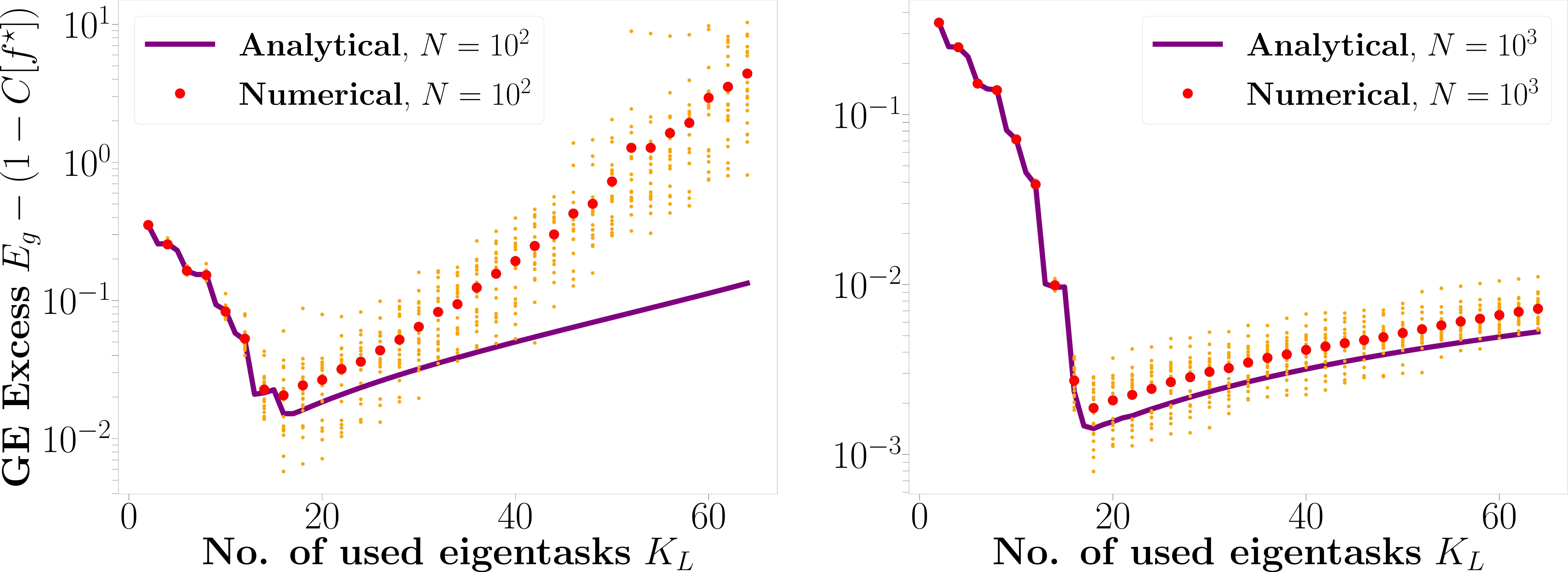}
    \caption{Generalization error excess $E_g - (1 -C[f^\star])$ vs number of used eigentasks $K_L$ for $S = 10^3$, where $K = 64$. The input data size are chosen to be $N=10^2$ (left) and $N=10^3$ (right). The task and Hamiltonian are the same as the ones in Fig.\,\ref{fig:1}, while the evolution time here is $\tau = 10$. 
    $R=20$ repetitions (in orange) are computed for each $K_L$. The mean values (red dots) and the theoretical error values (purple curve) are also compared. These two values match well at low $K_L$ for $N=10^2$ and fit all $K_L$ for $N=10^3$. This again shows the higher accuracy of replica theory for relatively large $N$.
    We also solved Eq.\,(\ref{eq:OTI_K}) and get the theoretical optimal truncation index $K_L^{\ast}(S)=15$ for $N=10^2$ and $K_L^{\ast}(S)=18$ for $N=10^3$. This is also consistent with the $K_L$, which corresponds to the minimal values of numerical generalization errors. }
    \label{fig:2}
\end{figure}

By increasing the number $K_L$, the prefactor in Eq.\,(\ref{eq:EgKL}) $N / \left( N - K_L - 1 \right)$ increases as well. It means this prefactor contributes to the generalization error as a characterization of overfitting. \rev{On the other hand, when including more eigentasks by increasing $K_L$, adding $y^{(k)}$ into the set of used eigentasks causes a change from $a_k^{2}$ to $ a_k^{2} \times \beta_k^2/(S + \beta_k^2) $ (namely, it causes from the first line to the second line in Eq.\,(\ref{eq:EgKL})).} Therefore, the term in parenthesis of Eq.\,(\ref{eq:EgKL}) decreases as $K_L$ increases. \rev{The different trends of the prefactor and the term in parenthesis of Eq.\,(\ref{eq:EgKL}) as we increase $K_L$ indicates an optimal $K_L$ which minimize $E_g$.}

To be more specific, if we evaluate the finite difference $\Delta E_g (K_L) = E_g (K_L + 1) - E_g (K_L)$, we will obtain
\begin{equation}
    \Delta E_g (K_L) \approx \frac{1 - C [f^{\star}]}{N (1 - \gamma)^2} - \frac{c_{K_L}^{2}}{(1 - \gamma) \left( 1 + \beta_{K_L}^2/S \right)} .
\end{equation}
The optimal value $K_L^{\ast}$ for $K_L$ should make $\Delta E_g (K_L)$ vanish. Such $K_L^{\ast}$ must be task-dependent. Namely, it depends on $\bm{a}$. However, the weights $\bm{a}$ are unknown in most practical settings. One way to estimate the \textit{optimal truncation index} $K_L^{\ast}$ is letting $a^{2}_k \approx \frac{1}{K}, \forall k \in [K]$ and then solving $\Delta E_g (K_L) = 0$. This approximation comes from the assumption that no eigentask $y^{(k)}$ should have a larger contribution than other eigentasks for a generic task $f^{\star}$. It gives equation of $K_L^{\ast}$
\begin{equation}
    \frac{\beta_{K_L^{\ast}}^2}{S} = \frac{N -K_L^{\ast}}{K - C_T (S)}, \label{eq:OTI_K}
\end{equation}
where Resolvable Expressive Capacity $C_T (S) = \sum_k 1/(1+\beta^2_k/S) \leq K$ quantifies the effective number of resolvable functions generated by quantum systems \cite{hu_tackling_2023}.
We remark here that as function of $K_L^{\ast}$, $\beta_{K_L^{\ast}}^2 / S$ is increasing while $(N -K_L^{\ast}) / (K - C_T (S))$ is decreasing. The intersection of these two curves yields the solution of the optimal truncation index. We also let $K_L^{\ast}= \min \{K, N\}$ if the above equation has no solution. Non-negativity of $\beta_{K_L^{\ast}}^2 / S > 0$ also ensures that the solution $K_L^{\ast} \leq N$.

In Fig.\,\ref{fig:2}, we present $E_g - (1 - C [f^{\star}])$, the deviation between the average generalization error $E_g$ and its large-$N$ limit $1 - C [f^{\star}]$, as a function of $K_L$. Examples are shown for two different input data sizes $N=10^2$ and $N=10^3$. 
The analytical and numerical generalization error not only matches well at small $K_L$ for the relatively small input size $N=10^2$, but also fits all $K_L$ perfectly for the relatively large input size $N=10^3$, this again indicates that replica theory presented in Sec.\,\ref{sec:Training_Generalization} works better for large dataset regime.
Also, by solving the equation of optimal truncation index Eq.\,(\ref{eq:OTI_K}), it gives a theoretical estimation $K_L^{\ast} (S) = 15$ and $K_L^{\ast} (S) = 18$, for $N=10^2$ and $10^3$ respectively. They also match the $K_L$, which minimizes the numerical generalization errors.
We emphasize here that by taking a look at the generalization errors in the left panel of Fig.\,\ref{fig:2}, we find applying Eigentask Learning order-truncation at $K_L = 15$ is $O(10^2)$ times better than keeping all features (namely $K_L = 64$). It presents that Eigentask Learning should be an essential procedure, especially when there is no sufficient data for learning in the small $N$ regime.

\section{Discussion}
\label{sec:discussion}

In this article, we formally study the training error and generalization error -- as a function of training dataset size $N$ and shot number $S$ in those QML schemes that use parametric quantum circuits without training the internal parameters. We use the replica method as the approximation trick to evaluate the training and generalization errors. The eventual expressions are in terms of the eigen-NSR, the important concept in eigentask analysis. The result can be immediately employed to provide a rigorous explanation of the efficacy of Eigentask Learning. 

\rev{However, several directions remain to be explored in the theory of Eigentask Learning. First of all, given the prior distribution $p(\bm{u})$ and encoding circuit $U(\bm{u})$, the distribution of spectrum $\{\beta^2_k\}$ is still not clear in general. This could cause difficulty in determining how many low-noise eigentask can be constructed from the quantum reservoir.}
\rev{Secondly, both the Eigentask Analysis and replica methods in Ref.\,\cite{canatar_spectral_2021} only consider the first- and second-order cumulants of input data (see also the Gaussian approximation in Appendix \ref{apx:GA}). Understanding the generalization error regarding higher-order correlation of data needs to be explored, and a new scope of feature representation beyond the Eigentask basis may be provided.}
Third, the analysis in this paper works when accurate values of $\bm{r}^{(k)}$ in the combination coefficients in eigentasks are known. In practice, those coefficients are unknown, and it remains a challenge to compute them as accurately as possible, especially those eigentasks with indices lower than threshold $K_L^{\ast}(S)$. 
Ref.\,\cite{hu_tackling_2023} provides a numerical method of estimating $\bm{r}^{(k)}$ from noisy experimental data by using singular value decomposition. The replica trick in this work is not easy to be generalized to compute numerical deviation of $\bm{r}^{(k)}$ because it is hard to use the weights expectations $\mathbb{E}_{\mathcal{D}}[w_k^\ast]$ and $\mathbb{E}_{\mathcal{D}}[w_j^\ast w_k^\ast]$ to express estimator of $\bm{r}^{(k)}$. Moreover, the numerical fluctuation of estimating $\bm{r}^{(k)}$ should also depend on the concrete methods used, case by case. But intuitively, those eigentasks with indices lower than threshold $K_L^{\ast}(S)$ are easier to obtain accurately since they are the features that are more noise-resilient. 

This presented work establishes a theoretical framework for leveraging fewer output features to fit a specified target function accurately. This advancement holds significant promise across various applications and has the potential to enhance the resource efficiency of numerous QML protocols \cite{innocenti_potential_2022, suzuki_natural_2022, hu_overcoming_2024, havlicek_supervised_2019}.

\section*{Acknowledgement}
F.\,H.\, would like to thank Hakan E. T\"ureci for stimulating discussions about the work that went into this manuscript. 

\appendix

\bibliography{bibtex}


\begin{widetext}
    \newpage





\startcontents[appendices]
\printcontents[appendices]{l}{1}{\section*{Appendices}\setcounter{tocdepth}{2}}

\makeatletter
\let\toc@pre\relax
\let\toc@post\relax
\makeatother

\newpage

\section{Definition of Generalization Error Function}
\label{apx:Generalization_Function}

In this section, we are going to give an detailed explanation about why Eq.\,(\ref{eq:def_epsilon_g}) is a valid definition of generalization error function for given weights $\bm{w}$. It will be helpful to first recall the loss function of any
input $\bm{u}$ and weights $\bm{w}$ is
\begin{equation}
    \mathcal{L} (\bm{w}, \mathcal{X} (\bm{u})) = \left( \bm{w} \cdot \left( \bm{x} (\bm{u}) + \frac{1}{\sqrt{S}} \bm{\zeta} (\bm{u}) \right) - \bm{c} \cdot \bm{x} (\bm{u}) - f_{\bot} (\bm{u}) \right)^2 .
\end{equation}
The following conditional expectation of loss function over the output samples $\mathcal{X} (\bm{u})$ (conditioned on the input $\bm{u}$) will be used frequently later
\begin{equation}
    \mathbb{E}_{\mathcal{X} (\bm{u})} [\mathcal{L} (\bm{w}, \mathcal{X} (\bm{u}))] = (\bm{w} - \bm{c})^T \bm{x} (\bm{u}) \bm{x} (\bm{u})^T (\bm{w} - \bm{c}) + \frac{1}{S} \bm{w}^T \mathbf{\Sigma} (\bm{u}) \bm{w} + f^2_{\bot} (\bm{u}) - 2 f_{\bot} (\bm{u}) \bm{x} (\bm{u})^T (\bm{w} - \bm{c}) . \label{eq:ApxA_lemma}
\end{equation}

As what we pointed in Sec.\,\ref{sec:Training_Generalization} of main text, generalization error $\epsilon_g$ can be understood in the scenario when quantum system sees a new, unseen testing dataset $\mathcal{D}' = \{ (\bm{v}^{(n')}, \mathcal{X} (\bm{v}^{(n')})) \}_{n' \in [N']}$. The testing error, which evaluates the performance of a \textit{fixed} $\bm{w}$, should be $\epsilon_{\mathrm{ts}} (\bm{w}) = \frac{1}{N'} \sum_{n' = 1}^{N'}(\bm{w} \cdot \bm{X} (\bm{v}^{(n')}) - f^{\star} (\bm{v}^{(n')}))^2$. Or more explicitly,
\begin{align}
    \epsilon_{\mathrm{ts}} (\bm{w}) = \frac{1}{N'} \sum_{n' = 1}^{N'} \mathcal{L} (\bm{w}, \mathcal{X} (\bm{v}^{(n')})) = \frac{1}{N'} \sum_{n' = 1}^{N'} \left( \bm{w} \cdot \left( \bm{x} (\bm{v}^{(n')}) + \frac{1}{\sqrt{S}} \bm{\zeta} (\bm{v}^{(n')}) \right) - (\bm{c} \cdot \bm{x} (\bm{v}^{(n')}) + f_{\bot} (\bm{v}^{(n')})) \right)^2 \label{eq:ets}
\end{align}
Now we show that $\epsilon_{\mathrm{ts}} (\bm{w})$ in Eq.\,(\ref{eq:ets}) should be close to $\mathbb{E}_{\bm{u}} [\mathbb{E}_{\mathcal{X} (\bm{u})} [\mathcal{L} (\bm{w}, \mathcal{X} (u))]]$ with arbitrarily high probability when $N' \rightarrow \infty$. This can be justified by two steps of argument. 

The first step is using the so-called \textit{concentration principle}. If $n_1' \neq n_2'$, then $u^{(n'_1)}$ and $u^{(n'_2)}$ are independent. Central limit theorem ensures that $\epsilon_{\rm{ts}}[\bm{w}^\ast]$ should be close to its expectation
\begin{align}
    & \mathbb{E}[\epsilon_{\mathrm{ts}}(\bm{w})] = \frac{1}{N'} \sum_{n' = 1}^{N'} \mathbb{E}_{\mathcal{X} (\bm{v}^{(n')})} \left[ \left( \bm{w} \cdot \left( \bm{x} (\bm{v}^{(n')}) + \frac{1}{\sqrt{S}} \bm{\zeta} (\bm{v}^{(n')}) \right) - (\bm{c} \cdot \bm{x} (\bm{v}^{(n')}) + f_{\bot} (\bm{v}^{(n')})) \right)^2 \right] \nonumber\\
    =~& \frac{1}{N'} \sum_{n' = 1}^{N'} (\bm{w} - \bm{c})^T \bm{x} (\bm{v}^{(n')}) \bm{x} (\bm{v}^{(n')})^T (\bm{w} - \bm{c}) + \frac{1}{S} \bm{w}^T \mathbf{\Sigma} (\bm{v}^{(n')}) \bm{w} + f^2_{\bot} (\bm{v}^{(n')}) - 2 f_{\bot} (\bm{v}^{(n')}) \bm{x} (\bm{v}^{(n')})^T (\bm{w} - \bm{c})  \label{eq:conineq}
\end{align}
with arbitrarily high probability when $N' \to \infty$, where the second line comes from Eq.\,(\ref{eq:ApxA_lemma}). This can shown formally by Hoeffding's inequality. Since both $\bm{w}$ and $f^{\star}$ are fixed, and the noisy features $\bm{X}(\bm{v}^{(n')}) = \bm{x} (\bm{v}^{(n')}) + \frac{1}{\sqrt{S}} \bm{\zeta} (\bm{v}^{(n')}) \in [0, 1]$. We just let $|\bm{w} \cdot \bm{X} (\bm{u}) - f^{\star} (\bm{u}))^2| < B$ for all $\bm{u}$, where $B$ is a positive constant. Then Hoeffding's inequality states for any $\varepsilon > 0$,
\begin{equation}
    \mathrm{Pr} \big[ \epsilon_{\mathrm{ts}}(\bm{w}) - \mathbb{E}[|\epsilon_{\mathrm{ts}}(\bm{w})]| \leq \varepsilon \big] \leq 2 \mathrm{exp} \left(-\frac{N'\varepsilon^2}{2B^2} \right),
\end{equation}
which proves that when $N' \to \infty$, testing error $\epsilon_{\mathrm{ts}}(\bm{w})$ will converge to its expectation $\mathbb{E}[\epsilon_{\mathrm{ts}}(\bm{w})]$ with arbitrarily high probability, even features $\bm{X}(\bm{v}^{(n')})$ are influenced by sampling noise.   

The second step is taking \textit{integration by Monte Carlo}: it ensures the summation Eq.\,(\ref{eq:conineq}) and the following integral
\begin{align}
    & \mathbb{E}_{\bm{u}} \left[ (\bm{w} - \bm{c})^T \bm{x} (\bm{u}) \bm{x} (\bm{u})^T (\bm{w} - \bm{c}) + \frac{1}{S} \bm{w}^T \mathbf{\Sigma} (\bm{u}) \bm{w} + f^2_{\bot} (\bm{u}) - 2 f_{\bot} (\bm{u}) \bm{x} (\bm{u})^T (\bm{w} - \bm{c}) \right] \nonumber\\
    =~& \bm{w}^T \left( \mathbf{G}+ \frac{1}{S} \mathbf{V} \right) \bm{w} - 2 \bm{c}^{T} \mathbf{G} \bm{w} + 1. 
\end{align}
should be close with arbitrarily high probability when $N' \to \infty$, where the last equality employs the definition of $\mathbf{G}=\mathbb{E}_{\bm{u}} [\bm{x} (\bm{u}) \bm{x} (\bm{u})^T]$, the definition of $\mathbf{V}=\mathbb{E}_{\bm{u}} [\mathbf{\Sigma} (\bm{u})]$, the orthogonality of $\int f_{\bot} (\bm{u}) \bm{x} (\bm{u}) \dd \bm{u} = 0$, and the normalization condition $\mathbb{E}_{\bm{u}} [f^{\star 2}] = \bm{c}^{T} \mathbf{G} \bm{c} +\mathbb{E}_{\bm{u}} [f^2_{\bot}] = 1$.

Therefore, we verify that when taking the large testing dataset size limitation $N' \to \infty$, the testing error should converge to a deterministic value
\begin{equation}
    \epsilon_g (\bm{w}) = \lim_{N' \rightarrow \infty} \epsilon_{\mathrm{ts}} (\bm{w}) = \bm{w}^T \left( \mathbf{G}+ \frac{1}{S} \mathbf{V} \right) \bm{w} - 2 \bm{c}^{T} \mathbf{G} \bm{w} + 1
\end{equation}
with probability $1$. We use this deterministic quantity as the definition of generalization function in Eq.\,(\ref{eq:def_epsilon_g}).

\section{Replica Theory for Solving Training and Generalization Errors}
\label{apx:Replica_Theory}

From now on, the remaining parts of Appendices is using replica theory to prove average training and generalization errors in Eq.\,(\ref{eq:Et}) and Eq.\,(\ref{eq:Eg}). Roughly speaking, replica tricks is a tool of calculating a physical quantity in one system, by taking $m$-replica duplication and then extrapolating to $m = 0$. 

Appendix \ref{apx:Replica_Theory} introduces the techniques of replica methods on analyzing the training and generalization errors. Then main result of this section is the replicated partition function $\mathbb{E}_{\mathcal{D}} [ Z ^m]$ in Eq.\,(\ref{eq:EDZm}). However, for solving the complicated Eq.\,(\ref{eq:EDZm}), we need to introduce the replica-symmetry ansatz and saddle point approximation, which are delivered in Appendix \ref{apx:Replica_Symmetry}. It reduces the computation of Eq.\,(\ref{eq:EDZm}) to $\mathbb{E}_{\mathcal{D}} [\ln Z]$ in Eq.\,(\ref{eq:RSFE}). Eventually, we reach the final expression of average training and generalization errors under basis of original features and eigentasks, in Appendix \ref{apx:Original_Features} and \ref{apx:Eigentask} respectively. The are exactly the Eq.\,(\ref{eq:Et}) and Eq.\,(\ref{eq:Eg}) presented in the main text. 

\subsection{Introducing Gibbs distribution}

Recall that the average training and generalization errors are $E_t =\mathbb{E}_{\mathcal{D}} [\epsilon_t (\bm{w}^{\ast} )]$ and $E_g =\mathbb{E}_{\mathcal{D}} [\epsilon_g (\bm{w}^{\ast} )]$, where
\begin{equation}
    \bm{w}^{\ast}  = \underset{\bm{w} \in \mathbb{R}^{K}}{\mathrm{argmin}} H (\bm{w}) = \underset{\bm{w} \in \mathbb{R}^{K}}{\mathrm{argmin}} \frac{1}{2 \lambda} \sum_{n = 1}^N \mathcal{L} (\bm{w}, \mathcal{X} (\bm{u}^{(n)})) + \frac{1}{2} \| \bm{w} \|^2 . \label{eq:wD_def}
\end{equation}
In principle, the optimized weights $\bm{w}^{\ast} $ has a closed algebraic form by solving Eq.\,(\ref{eq:wD_def}). However, it is still hard to compute the quenched disorder ensemble average over such random $\bm{w}^{\ast} $, which is complicatedly $\mathcal{D}$-dependent. One solution is introducing the Gibbs distribution
\begin{equation}
    p_G (\bm{w}, \beta) = \frac{e^{- \beta H (\bm{w}  ) }}{Z [\beta ] } ,
\end{equation}
where the partition function $Z [\beta  ] = \int \dd \bm{w} e^{- \beta H (\bm{w}  ) }$, and then take $\beta \to \infty$ limit:
\begin{equation}
    p_G (\bm{w}, \beta \rightarrow \infty  ) = \delta (\bm{w} - \bm{w}^{\ast} ) .
\end{equation}
This trick comes from the observation that the stationary distribution of a standard stochastic gradient descent is Gibbs distribution, whose low temperature limit corresponds to the stochasticity-vanishing situation \cite{seung_statistical_1992}. Then, for any function $f$ we have
\begin{equation}
    \mathbb{E}_{\mathcal{D}} [f (\bm{w}^{\ast} )] = \lim_{\beta \rightarrow \infty} \mathbb{E}_{\mathcal{D}} \left[ \int \dd \bm{w} p_G (\bm{w}, \beta \rightarrow \infty  ) f (\bm{w} ) \right] .
\end{equation}

\subsection{Adding source terms}

By observing average generalization error $E_g =\mathbb{E}_{\mathcal{D}} [\epsilon_g (\bm{w}^{\ast})] =\mathbb{E}_{\mathcal{D}} \left[ \bm{w}^{\ast T} \left( \mathbf{G}+ \frac{1}{S} \mathbf{V} \right) \bm{w}^{\ast} - 2 \bm{c}^{T} \mathbf{G} \bm{w}^{\ast} \right] + 1$, we find that as long as all terms like $\mathbb{E}_{\mathcal{D}} [w_k^{\ast}]$ and $\mathbb{E}_{\mathcal{D}} [w_j^{\ast} w_k^{\ast}]$ are known, the average generalization error can be computed by taking the linear combination
\begin{equation}
    E_g = \sum_{j, k = 1}^K \left( \mathbf{G}_{j k} + \frac{1}{S} \mathbf{V}_{j k} \right) \mathbb{E}_{\mathcal{D}} [w_j^{\ast} w_k^{\ast}] - 2 \sum_{j, k = 1}^K c_j \mathbf{G}_{j k} \mathbb{E}_{\mathcal{D}} [w_k^{\ast}] + 1.
\end{equation}
Just as what is usually done in quantum field theory and statistical mechanics, such first and second order moment can be obtained by adding source terms in partition function. Taking derivative near the zero-source yield those moments. Hence, it suggests us to add the source term in a newly defined partition function:
\begin{equation}
    Z [\bm{\xi}, \bm{\eta}, \beta ] = \int \dd \bm{w} e^{- \beta H (\bm{w} ) + \beta \bm{\xi}^T \bm{w} + \frac{1}{2} \beta \bm{\eta}^T \bm{w}  \bm{w}^T \bm{\eta}} .
\end{equation}
such that
\begin{align}
    \mathbb{E}_{\mathcal{D}} [w_k^{\ast}] & = \lim_{\beta \rightarrow \infty} \frac{1}{\beta} \frac{\partial}{\partial \xi_k} \mathbb{E}_{\mathcal{D}} [\ln Z [\bm{\xi}, \bm{\eta}, \beta ]] |_{\bm{\xi}, \bm{\eta} = \bm{0}}, \label{eq:EDwast-1-old} \\
    \mathbb{E}_{\mathcal{D}} [ w_j^{\ast} w_k^{\ast} ] & = \lim_{\beta \rightarrow \infty} \left. \frac{1}{\beta} \frac{\partial^2}{\partial \eta_j \partial \eta_k} \mathbb{E}_{\mathcal{D}} [\ln Z [\bm{\xi}, \bm{\eta}, \beta ]] \right|_{\bm{\xi}, \bm{\eta} = \bm{0}} . \label{eq:EDwast-2-old}
\end{align}
Furthermore, we can verify that the new partition function $Z [\bm{\xi}, \bm{\eta}, \beta ]$ can also generate the average training error Eq.\,(\ref{eq:def_epsilon_t}):
\begin{align}
    E_t & = \frac{1}{N} \sum_{n = 1}^N \mathbb{E}_{\mathcal{D}} \left[ \left( (\bm{w}^{\ast} - \bm{c}) \cdot \bm{x} (\bm{u}^{(n)}) + \frac{1}{\sqrt{S}} \bm{w}^{\ast} \cdot \bm{\zeta} (\bm{u}^{(n)}) - f_{\bot} (\bm{u}^{(n)}) \right)^2 \right] = \frac{2 \lambda}{N} \left( \mathbb{E}_{\mathcal{D}} [H (\bm{w}^{\ast} )] - \frac{1}{2} \mathbb{E}_{\mathcal{D}} [\| \bm{w}^{\ast} \|^2] \right) \nonumber\\
    & = \frac{2 \lambda}{N} \left( - \lim_{\beta \rightarrow \infty} \frac{\partial}{\partial \beta} \mathbb{E}_{\mathcal{D}} [\ln Z [\bm{\xi}, \bm{\eta}, \beta ] |_{\bm{\xi}, \bm{\eta} = \bm{0}}] - \frac{1}{2} \mathbb{E}_{\mathcal{D}} [\| \bm{w}^{\ast} \|^2] \right) . 
\end{align}

\subsection{Replica trick}
\label{apx:Replica_trick}

Replica trick is a commonly used trick for dealing with quantity involving logarithm, in both high energy theory and machine learning theory. The mathematical intuition is the limitation identity
\begin{equation}
    \ln Z = \lim_{m \rightarrow 0} \frac{Z^m - 1}{m} . \label{eq:relica_lnZ}
\end{equation}
By taking expectation over $\mathcal{D}$ on both sides of Eq.\,(\ref{eq:relica_lnZ}), we can rewrite the averaged free energy into a limitation form $\mathbb{E}_{\mathcal{D}} [\ln Z] = \lim_{m \rightarrow 0} \frac{\mathbb{E}_{\mathcal{D}} [Z^m] - 1}{m}$.

Suppose a system with $m \in \mathbb{N}^+$ replicas of the original quantum systems, which have $m$ different weights $\{ \bm{w}^{\mu} \}_{\mu \in [m]}$ but share the same quenched disorder $\mathcal{D}$:
\begin{align}
    Z^m [\bm{\xi}, \bm{\eta}, \beta ] & = \int \left\{ \prod_{\mu = 1}^m \dd \bm{w}^{\mu} \right\} e^{- \beta \sum^m_{\mu = 1} \left( H (\bm{w}^{\mu} ) + \beta \bm{\xi}^T \bm{w}^{\mu} + \frac{\beta}{2} \bm{\eta}^T \bm{w}^{\mu} \bm{w}^{\mu T} \bm{\eta} \right)} \nonumber\\
    & = \int \left\{ \prod_{\mu = 1}^m \dd \bm{w}^{\mu} \right\}  e^{- \beta \sum^m_{\mu = 1} \left( \frac{1}{2} \bm{w}^{\mu T} \bm{w}^{\mu} - \bm{\xi}^T \bm{w}^{\mu} - \frac{1}{2} \bm{\eta}^T \bm{w}^{\mu} \bm{w}^{\mu T} \bm{\eta} \right)} \nonumber\\
    & \hspace{7.5em} \times e^{- \beta \sum^m_{\mu = 1} \sum^N_{n = 1} \left( (\bm{w}^{\mu} - \bm{c}) \cdot \bm{x} (\bm{u}^{(n)}) + \frac{1}{\sqrt{S}} \bm{w}^{\mu} \cdot \bm{\zeta} (\bm{u}^{(n)}) - f_{\bot} (\bm{u}^{(n)}) \right)^2} . 
\end{align}
By taking the transformation $\bm{w}^{\mu} \rightarrow \bm{w}^{\mu} + \bm{c}$ and define $\bm{W}^{\star} = (\bm{c} - \bm{\xi}) - (\bm{\eta}^T \bm{c}) \bm{\eta}$, one get
\begin{align}
    & Z^m [\bm{\xi}, \bm{\eta}, \beta ] \nonumber\\
    =~& e^{- \frac{m \beta}{2} (\bm{W}^{\star} - \bm{\xi})^T \bm{c}} \int \left\{ \prod_{\mu = 1}^m \dd \bm{w}^{\mu} \right\} e^{- \frac{\beta}{2} \sum^m_{\mu = 1} \bm{w}^{\mu T} (\mathbf{I}- \bm{\eta} \bm{\eta}^T) \bm{w}^{\mu} - \beta \bm{W}^{\star T} \sum^m_{\mu = 1} \bm{w}^{\mu}} \nonumber\\
    & \hspace{13.5em} \times e^{- \frac{\beta}{2 \lambda} \sum^m_{\mu = 1} \sum^N_{n = 1} \left( \bm{w}^{\mu} \cdot \bm{x} (\bm{u}^{(n)}) + \frac{1}{\sqrt{S}} (\bm{w}^{\mu} + \bm{c}) \cdot \bm{\zeta} (\bm{u}^{(n)}) - f_{\bot} (\bm{u}^{(n)}) \right)^2} . \label{eq:Zm_int_1}
\end{align}
The first term $e^{- \frac{\beta}{2} \sum^m_{\mu = 1} \bm{w}^{\mu T} (\mathbf{I}- \bm{\eta} \bm{\eta}^T) \bm{w}^{\mu} - \beta \bm{W}^{\star T} \sum^m_{\mu = 1} \bm{w}^{\mu}}$ is independent of the quenched disorder $\mathcal{D}$. The quenched disorder ensemble average of the second term can be simplified into form of exponential due to the fact that $\mathcal{I}= \{ \bm{u}^{(n)} \}$ are i.i.d., and $\mathcal{X} (\bm{u}^{(n_1)})$ and $\mathcal{X} (\bm{u}^{(n_2)})$ should be independent if $n_1 \neq n_2$:
\begin{align}
    & \mathbb{E}_{\mathcal{D}} \left[ e^{- \frac{\beta}{2 \lambda} \sum^m_{\mu = 1} \sum^N_{n = 1} \left( \bm{w}^{\mu} \cdot \bm{x} (\bm{u}^{(n)}) + \frac{1}{\sqrt{S}} (\bm{w}^{\mu} + \bm{c}) \cdot \bm{\zeta} (\bm{u}^{(n)}) - f_{\bot} (\bm{u}^{(n)}) \right)^2} \right] \nonumber\\
    =~& \mathbb{E}_{\mathcal{D}} \left[ e^{- \frac{\beta}{2 \lambda} \sum^m_{\mu = 1} \left( \bm{w}^{\mu} \cdot \bm{x} (\bm{u}) + \frac{1}{\sqrt{S}} (\bm{w}^{\mu} + \bm{c}) \cdot \bm{\zeta} (\bm{u}) - f_{\bot} (\bm{u}) \right)^2} \right]^N
\end{align}
where
\begin{align}
    \mathbb{E}_{\mathcal{D}} \!\! \left[ e^{- \frac{\beta}{2 \lambda} \sum^m_{\mu = 1} \left( \bm{w}^{\mu} \cdot \bm{x} (\bm{u}) + \frac{1}{\sqrt{S}} (\bm{w}^{\mu} + \bm{c}) \cdot \bm{\zeta} (\bm{u}) - f_{\bot} (\bm{u}) \right)^2} \right] = \mathbb{E}_{\bm{u}} \!\! \left[ \mathbb{E}_{\mathcal{X} (\bm{u})} \!\! \left[ e^{- \frac{\beta}{2 \lambda} \sum^m_{\mu = 1} \left( \bm{w}^{\mu} \cdot \bm{x} (\bm{u}) + \frac{1}{\sqrt{S}} (\bm{w}^{\mu} + \bm{c}) \cdot \bm{\zeta} (\bm{u}) - f_{\bot} (\bm{u}) \right)^2} \right] \right]. \label{eq:Zm_int_2}
\end{align}
There are two difficult integrals in Eq.\,(\ref{eq:Zm_int_1}). The first one is the integral with respect to quenched disorder in Eq.\,(\ref{eq:Zm_int_2}); the second one is the integral with respect to the replica weights $\left\{ \prod_{\mu = 1}^m \dd \bm{w}^{\mu} \right\}$ in Eq.\,(\ref{eq:Zm_int_1}). We need to tackle them separately. 

\subsection{Integrate with respect to quenched disorder}
\label{apx:GA}
The expectation in Eq.\,(\ref{eq:Zm_int_2}) is still general and does not has a closed form. By looking at the quadratic term at the exponent in Eq.\,(\ref{eq:Zm_int_1}), we adopt the Gaussian approximation used in Ref.\,\cite{canatar_statistical_2022}, and define $m$ random variables $q^{\mu} = \bm{w}^{\mu} \cdot \bm{x} (\bm{u}) + \frac{1}{\sqrt{S}} (\bm{w}^{\mu} + \bm{c}) \cdot \bm{\zeta} (\bm{u}) - f_{\bot} (\bm{u})$, whose mean $\bm{R}$ and
covariance $\bm{Q}$ are
\begin{align}
    R^{\mu} := \mathbb{E}_{\mathcal{D}} [q^{\mu}] & = \mathbb{E}_{\bm{u}} [\bm{w}^{\mu T} \bm{x}] + \frac{1}{\sqrt{S}} \mathbb{E}_{\bm{u}} [ \mathbb{E}_{\mathcal{X} (\bm{u})} [ (\bm{w}^{\mu} + \bm{c})^T \bm{\zeta} ] ] - \mathbb{E}_{\bm{u}} [ f_{\bot}] = \bm{w}^{\mu T} \bm{d}, \\
    Q^{\mu \nu} := \mathbb{E}_{\mathcal{D}} [q^{\mu} q^{\nu}] - \mathbb{E}_{\mathcal{D}} [q^{\mu}] \mathbb{E}_{\mathcal{D}} [q^{\nu}] & = \mathbb{E}_{\bm{u}} [\bm{w}^{\mu T} \bm{x} \bm{x}^T \bm{w}^{\nu}] + \frac{1}{S} \mathbb{E}_{\bm{u}} [ \mathbb{E}_{\mathcal{X} (\bm{u})} [(\bm{w}^{\mu} + \bm{c})^T \bm{\zeta} \bm{\zeta}^T (\bm{w}^{\nu} + \bm{c}) ] ] + \mathbb{E}_{\bm{u}} [ f_{\bot}^2] - \bm{w}^{\mu T} \bm{d} \bm{d}^T \bm{w}^{\nu} \nonumber\\
    & = \bm{w}^{\mu T} \mathbf{G} \bm{w}^{\nu} + \frac{1}{S} (\bm{w}^{\mu} + \bm{c})^T \mathbf{V} (\bm{w}^{\nu} + \bm{c}) + \mathbb{E}_{\bm{u}} [ f_{\bot}^2] - \bm{w}^{\mu T} \bm{d} \bm{d}^T \bm{w}^{\nu} \nonumber \\
    & = \bm{w}^{\mu T} \left( \mathbf{G}+ \frac{1}{S} \mathbf{V}- \bm{d} \bm{d}^T \right) \bm{w}^{\nu} + \frac{1}{S} (\bm{w}^{\mu} + \bm{w}^{\nu})^T \mathbf{V} \bm{c} + \frac{1}{S} \bm{c}^{T} \mathbf{V} \bm{c} +\mathbb{E}_{\bm{u}} [f^2_{\bot}] , 
\end{align}
where we use $\mathbb{E}_{\mathcal{X} (\bm{u})} [ \bm{\zeta} ] = 0$, $\mathbb{E}_{\bm{u}} [ f_{\bot}] = \sum_k \mathbb{E}_{\bm{u}} [ x_k f_{\bot} ] = 0$ and define $d_k = \int \dd u p (u) x_k (u)$. Notice that the integrated second-order-moment matrix $\mathbf{D} := \mathbf{V}-\mathbf{G}$ studied in Ref.\,\cite{hu_tackling_2023} is a diagonal matrix with exactly $\mathbf{D}_{k k} = d_k$ for all $k$. Also, for convenience, we introduce two new matrices
\begin{align}
    \mathbf{C} & = \mathbf{G}+ \frac{1}{S} \mathbf{V} - \bm{d} \bm{d}^T . \\
    \widetilde{\mathbf{\Sigma}} & = \tilde{\sigma}^2 \mathbf{1} \mathbf{1}^T \quad \mathrm{where} \quad \tilde{\sigma}^2 = \frac{1}{S} \bm{c}^{T} \mathbf{V} \bm{c} +\mathbb{E}_{\bm{u}} [f^2_{\bot}].
\end{align}

We now introduce a Gaussian approximation $\bm{q} \sim \mathcal{N}(\bm{R}, \bm{Q})$ here. It assumes $\bm{q} \in \mathbb{R}^m$ is approximately a Gaussian random variable with mean $\bm{R}$ and covariance $\bm{Q}$, which are computed above. The discussion of the validity of the Gaussian approximation can be found in Ref.\,\cite{canatar_spectral_2021}.
This approximation is aligned with the same idea in Ref.\,\cite{hu_tackling_2023} where one ignores the contribution of high-order cumulants of features, and only takes the first and second-order cumulants of features into account.
Under this Gaussian approximation, the distribution of $\bm{q}$ under Gaussian approximation is explicitly
\begin{equation}
    P (\bm{q}) = \frac{1}{(2 \pi)^{\frac{m}{2}} \det (\bm{Q})^{\frac{1}{2}}} e^{- \frac{1}{2} (\bm{q} - \bm{R})^T \bm{Q}^{- 1} (\bm{q} - \bm{R})} .
\end{equation}
Therefore, the integral in Eq.\,(\ref{eq:Zm_int_2}) is
\begin{align}
    & \mathbb{E}_{\mathcal{D}} \left[ e^{- \frac{\beta}{2 \lambda} \sum^m_{\mu = 1} \left( \bm{w}^{\mu} \cdot \bm{x} (\bm{u}) + \frac{1}{\sqrt{S}} (\bm{w}^{\mu} + \bm{c}) \cdot \bm{\zeta} (\bm{u}) - f_{\bot} (\bm{u}) \right)^2} \right] \approx \int \dd \bm{q} \frac{1}{(2 \pi)^{\frac{m}{2}} \det (\bm{Q})^{\frac{1}{2}}} e^{- \frac{1}{2} (\bm{q} - \bm{R})^T \bm{Q}^{- 1} (\bm{q} - \bm{R}) - \frac{\beta}{2 \lambda} \bm{q}^T \bm{q}} \nonumber\\
    =~& \det \left( \mathbf{I}+ \frac{\beta}{\lambda} \bm{Q} \right)^{- \frac{1}{2}} e^{- \frac{\beta}{2 \lambda} \bm{R}^T \left( \mathbf{I}+ \frac{\beta}{\lambda} \bm{Q} \right)^{- 1} \bm{R}}, 
\end{align}
where we have used identity $\left( \mathbf{I}+ \frac{\beta}{\lambda} \bm{Q} \right)^{- 1} =\mathbf{I}- \left( \mathbf{I}+ \frac{\beta}{\lambda} \bm{Q} \right)^{- 1} \frac{\beta}{\lambda} \bm{Q}$ to simplify.

\subsection{Integrate with respect to replica weights}

We have already integrated with respect to quenched disorder $\mathcal{D}$, the cost is that we have to introduce two new variables $\bm{R}$ and $\bm{Q}$, which are highly complicated functions of $\bm{w}$. Then the integral with respect to $\{ \prod_{\mu} \dd \bm{w}^{\mu} \}$ in Eq.\,(\ref{eq:Zm_int_1}) is still hard. According to the identity of Dirac function $\delta (a - \alpha) = \frac{1}{2 \pi} \int \dd \hat{a} e^{i \hat{a} (a - \alpha)} = \frac{1}{2 \pi} \int \dd (i_{} N \hat{a}) e^{i (i_{} N \hat{a}) (a - \alpha)} = \frac{i_{} N}{2 \pi} \int \dd \hat{a} e^{- N \hat{a} (a - \alpha)}$, we can insert the following two Dirac functions
\begin{align}
    \delta (R^{\mu} - \bm{w}^{\mu T} \bm{d}) & = \frac{i N}{2 \pi} \int \dd \hat{R}^{\mu} e^{- N \hat{R}^{\mu} (R^{\mu} - \bm{w}^{\mu T} \bm{d})}, \\ 
    \delta \left( Q^{\mu \nu} - \bm{w}^{\mu T} \mathbf{C} \bm{w}^{\nu} - \frac{1}{S} (\bm{w}^{\mu} + \bm{w}^{\nu})^T \mathbf{V} \bm{c} - \tilde{\sigma}^2 \right) & = \frac{i N}{2 \pi} \int \dd \hat{Q}^{\mu \nu} e^{- N \hat{Q}^{\mu \nu} \left( Q^{\mu \nu} - \bm{w}^{\mu T} \mathbf{C} \bm{w}^{\nu} - \frac{1}{S} (\bm{w}^{\mu} + \bm{w}^{\nu})^T \mathbf{V} \bm{c} - \tilde{\sigma}^2 \right)}, 
\end{align}
such that
\begin{align}
    & \mathbb{E}_{\mathcal{D}} [Z^m [\bm{\xi}, \bm{\eta}, \beta ]] \nonumber\\
    =~& e^{- \frac{m \beta}{2} (\bm{W}^{\star} - \bm{\xi})^T \bm{c}} \int \left\{ \prod_{\mu = 1}^m \dd \bm{w}^{\mu} \right\} e^{- \beta \sum^m_{\mu = 1} \left( \frac{1}{2} \bm{w}^{\mu T} (\mathbf{I}- \bm{\eta} \bm{\eta}^T) \bm{w}^{\mu} + \bm{W}^{\star T} \bm{w}^{\mu} \right)} e^{- N \left( \frac{1}{2} \ln \det \left( \mathbf{I}+ \frac{\beta}{\lambda} \bm{Q} \right) + \frac{\beta}{2 \lambda} \bm{R}^T \left( \mathbf{I}+ \frac{\beta}{\lambda} \bm{Q} \right)^{- 1} \bm{R} \right)} \nonumber\\
    =~& e^{- \frac{m \beta}{2} (\bm{W}^{\star} - \bm{\xi})^T \bm{c}} \int \left\{ \prod_{\mu} \dd R^{\mu} \prod_{\mu \geq \nu} \dd Q^{\mu \nu} \right\} e^{- N \left( \frac{1}{2} \ln \det \left( \mathbf{I}+ \frac{\beta}{\lambda} \bm{Q} \right) + \frac{\beta}{2 \lambda} \bm{R}^T \left( \mathbf{I}+ \frac{\beta}{\lambda} \bm{Q} \right)^{- 1} \bm{R} \right)} \nonumber\\
    & \!\!\times \!\! \int\! \left\{ \prod_{\mu = 1}^m \dd \bm{w}^{\mu} \right\} e^{- \beta \sum^m_{\mu = 1} \left( \frac{1}{2} \bm{w}^{\mu T} (\mathbf{I}- \bm{\eta} \bm{\eta}^T) \bm{w}^{\mu} + \bm{W}^{\star T} \bm{w}^{\mu} \right)} \delta (R^{\mu} - \bm{w}^{\mu T} \bm{d}) \delta \left( Q^{\mu \nu} - \bm{w}^{\mu T} \mathbf{C} \bm{w}^{\nu} - \frac{1}{S} (\bm{w}^{\mu} + \bm{w}^{\nu})^T \mathbf{V} \bm{c} - \tilde{\sigma}^2 \right) \nonumber\\
    =~& \left( \frac{i N}{2 \pi} \right)^{\frac{m (m + 3)}{2}} e^{- \frac{m \beta}{2} (\bm{W}^{\star} - \bm{\xi})^T \bm{c}} \int \left\{ \prod_{\mu} \dd R^{\mu} \dd \hat{R}^{\mu} \right\} \left\{ \prod_{\mu \geq \nu} \dd Q^{\mu \nu} \dd \hat{Q}^{\mu \nu} \right\} e^{- N \left( \frac{1}{2} \ln \det \left( \mathbf{I}+ \frac{\beta}{\lambda} \bm{Q} \right) + \frac{\beta}{2 \lambda} \bm{R}^T \left( \mathbf{I}+ \frac{\beta}{\lambda} \bm{Q} \right)^{- 1} \bm{R} \right)} \times \nonumber\\
    & \!\! \int \!\! \left\{ \prod_{\mu = 1}^m \dd \bm{w}^{\mu} \right\} e^{- \beta \sum_{\mu} \left( \frac{1}{2} \bm{w}^{\mu T} (\mathbf{I}- \bm{\eta} \bm{\eta}^T) \bm{w}^{\mu} + \bm{W}^{\star T} \bm{w}^{\mu} \right) - N \sum_{\mu} \hat{R}^{\mu} (R^{\mu} - \bm{w}^{\mu T} \bm{d}) - N \sum_{\mu \geq \nu} \hat{Q}^{\mu \nu} \left( Q^{\mu \nu} - \bm{w}^{\mu T} \mathbf{C} \bm{w}^{\nu} - \frac{1}{S} (\bm{w}^{\mu} + \bm{w}^{\nu})^T \mathbf{V} \bm{c} - \tilde{\sigma}^2 \right)} . 
\end{align}
The last line contains a Gaussian integral with form of $\int \dd \bm{w}^{\oplus} e^{- \frac{\beta}{2} \bm{w}^{\oplus T} \mathbf{X} \bm{w}^{\oplus} - \beta \bm{W}^{\oplus T} \bm{w}^{\oplus}}$ in $m K$ dimensional space with parameters
\begin{align}
    \bm{w}^{\oplus} & = \bigoplus_{\mu = 1}^m \bm{w}^{\mu}, \\
    \bm{W}^{\oplus} & = \bigoplus_{\mu = 1}^m \left( \bm{W}^{\star} - \frac{N}{\beta} \hat{R}^{\mu} \bm{d} - \frac{N}{\beta S} \left( \sum_{\nu = 1}^m \hat{Q}^{\mu \nu} (1 + \delta^{\mu \nu}) \right) \mathbf{V} \bm{c} \right), \\
    \mathbf{X} & = 
        \left(\begin{array}{cccc}
            \mathbf{X}^{11} & \mathbf{X}^{12} & \cdots & \mathbf{X}^{1 m}\\
            \mathbf{X}^{21} & \mathbf{X}^{22} & \cdots & \mathbf{X}^{2 m}\\
            \vdots & \vdots & \ddots & \vdots\\
            \mathbf{X}^{m 1} & \mathbf{X}^{m 2} & \cdots & \mathbf{X}^{m m}
        \end{array}\right), \\
    \mathbf{X}^{\mu \nu} & = \delta^{\mu \nu} (\mathbf{I}- \bm{\eta} \bm{\eta}^T) - \frac{N}{\beta} \hat{Q}^{\mu \nu} (1 + \delta^{\mu \nu}) \mathbf{C}. 
\end{align}
Finally, we get the replicated partition function, under quenched disorder ensemble average
\begin{align}
    \mathbb{E}_{\mathcal{D}} [Z^m [\bm{\xi}, \bm{\eta}, \beta ]] = e^{\frac{m (m + 3)}{2} \ln \left( \frac{i N}{2 \pi} \right) + \frac{m K}{2} \ln \left( \frac{2 \pi}{\beta} \right)} \int \left\{ \prod_{\mu} \dd R^{\mu} \dd \hat{R}^{\mu} \right\} \left\{ \prod_{\mu \leq \nu} \dd Q^{\mu \nu} \dd \hat{Q}^{\mu \nu} \right\} e^{- m N\mathcal{S} [\bm{Q}, \hat{\bm{Q}}, \bm{R}, \hat{\bm{R}}]},  \label{eq:EDZm}
\end{align}
where the \textit{free energy} $\mathcal{S}$ is
\begin{align}
    \mathcal{S} [\bm{Q}, \hat{\bm{Q}}, \bm{R}, \hat{\bm{R}}] & = \frac{1}{m} \sum_{\mu} \hat{R}^{\mu} R^{\mu} + \frac{1}{m} \sum_{\mu \geq \nu} \hat{Q}^{\mu \nu} (Q^{\mu \nu} - \widetilde{\mathbf{\Sigma}}^{\mu \nu}) + \frac{1}{m N} \left( \mathcal{G}_0 + N\mathcal{G}_r + \frac{m \beta}{2} (\bm{W}^{\star} - \bm{\xi})^T \bm{c} \right), \\
    \mathcal{G}_0 [\bm{Q}, \hat{\bm{Q}}, \bm{R}, \hat{\bm{R}}] & = \frac{1}{2} \ln \det (\mathbf{X}) - \frac{\beta}{2} \bm{W}^{\oplus T} \mathbf{X}^{- 1} \bm{W}^{\oplus}, \\
    \mathcal{G}_r [\bm{Q}, \bm{R}] & = \frac{1}{2} \ln \det \left( \mathbf{I}+ \frac{\beta}{\lambda} \bm{Q} \right) + \frac{\beta}{2 \lambda} \bm{R}^T \left( \mathbf{I}+ \frac{\beta}{\lambda} \bm{Q} \right)^{- 1} \bm{R} . 
\end{align}

\noindent \textbf{Remark}. In practice we always have $N \gg 1$, it means we can use \textit{saddle point approximation} to evaluate Eq.\,(\ref{eq:EDZm}). More explicitly, saddle points of exponent dominates the integral $\int \left\{ \prod_{\mu} \dd R^{\mu} \dd \hat{R}^{\mu} \right\} \left\{ \prod_{\mu \leq \nu} \dd Q^{\mu \nu} \dd \hat{Q}^{\mu \nu} \right\} e^{- m N\mathcal{S} [\bm{Q}, \hat{\bm{Q}}, \bm{R}, \hat{\bm{R}}]}$. Replica trick then extrapolates this integral value from $m \in \mathbb{N}^+$ to $m \in \mathbb{R}$, and take the limit $m \rightarrow 0$:
\begin{equation}
    \mathbb{E}_{\mathcal{D}} [\ln Z [\bm{\xi}, \bm{\eta}, \beta ]] = \lim_{m \rightarrow 0} \frac{\mathbb{E}_{\mathcal{D}} [Z^m [\bm{\xi}, \bm{\eta}, \beta ]] - 1}{m} = - N \underset{\bm{Q}, \hat{\bm{Q}}, \bm{R}, \hat{\bm{R}}}{\mathrm{extr}} \{ \mathcal{S} [\bm{Q}, \hat{\bm{Q}}, \bm{R}, \hat{\bm{R}}] \} . \label{eq:rtm0}
\end{equation}

\section{Replica Symmetry Solutions}
\label{apx:Replica_Symmetry}

\subsection{Replica symmetry ansatz}
\label{apx:rsa}

Now we have already transformed the calculation of $\mathcal{D}$-averaged log-partition function $\mathbb{E}_{\mathcal{D}} [\ln Z [\bm{\xi}, \bm{\eta}, \beta ]]$ into a minimization form of $\mathcal{S} [\bm{Q}, \hat{\bm{Q}}, \bm{R}, \hat{\bm{R}}]$. However, free energy function $\mathcal{S} [\bm{Q}, \hat{\bm{Q}}, \bm{R}, \hat{\bm{R}}]$ has $m^2 + 3 m$ degrees of freedom. A reasonable way to deal with it is assuming all its saddle points satisfy a certain symmetry, this is exactly the well-known \textit{replica symmetry ansatz} (RSA):
\begin{eqnarray}
    Q^{\mu \mu} = q_0, \quad & \hat{Q}^{\mu \mu} = \hat{q}_0, & \quad R^{\mu} = r, \nonumber\\ Q^{\mu \neq \nu} = q_1, \quad & \hat{Q}^{\mu \neq \nu} = \hat{q}_1, & \quad \hat{R}^{\mu} = \hat{r} . 
\end{eqnarray}
Thus, the parameters in $\mathcal{G}_0$ are simplified to
\begin{align}
    \bm{W}^{\oplus} & = \bigoplus_{\mu = 1}^m \bm{W}^{\star} - \frac{N \hat{r}}{\beta} \bm{d} - \frac{N}{\beta S} (2 \hat{q}_0 + (m - 1) \hat{q}_1) \mathbf{V} \bm{c}, \\
    \mathbf{X}_0 & := \mathbf{X}^{\mu \mu} = (\mathbf{I}- \bm{\eta} \bm{\eta}^T) - \frac{2 N \hat{q}_0}{\beta} \mathbf{C}, \\ 
    \mathbf{X}_1 & := \mathbf{X}^{\mu \neq \nu} = - \frac{N \hat{q}_1}{\beta} \mathbf{C}. 
\end{align}
Furthermore, by truncating the second and higher order terms of $m$ in these listed quantities:
The second and higher order of the listed quantities are truncated.
we use Eq.\,(\ref{eq:rtm0}) and get the \textit{replica-symmetric free energy} is
\begin{align}
    \mathcal{S} [Q, \hat{Q}, R, \hat{R}] =~& r \hat{r} + (q_0 - \tilde{\sigma}^2) \hat{q}_0 - \frac{1}{2} (q_1 - \tilde{\sigma}^2) \hat{q}_1 + \frac{1}{2} \ln \left( 1 + \frac{\beta}{\lambda} (q_0 - q_1) \right) + \frac{1}{2} \frac{\beta (q_1 + r^2)}{\lambda + \beta (q_0 - q_1)} \nonumber\\
    & + \frac{1}{2 N} \left( \ln \det (\bm{\Delta}) - \frac{N \hat{q}_1}{\beta} \mathrm{Tr} (\mathbf{C} \bm{\Delta}^{- 1}) + \beta (\bm{W}^{\star} - \bm{\xi})^T \bm{c} \right) \nonumber\\
    & - \frac{\beta}{2 N} \left( \bm{W}^{\star} - \frac{N \hat{r}}{\beta} \bm{d} - \frac{N (2 \hat{q}_0 - \hat{q}_1)}{\beta S} \mathbf{V} \bm{c} \right)^T \bm{\Delta}^{- 1} \left( \bm{W}^{\star} - \frac{N \hat{r}}{\beta} \bm{d} - \frac{N (2 \hat{q}_0 - \hat{q}_1)}{\beta S} \mathbf{V} \bm{c} \right), 
\end{align}
where we recall $\bm{\Delta} = (\mathbf{I}- \bm{\eta} \bm{\eta}^T) - \frac{N (2 \hat{q}_0 - \hat{q}_1)}{\beta} \mathbf{C}$ and $\bm{W}^{\star} = (\bm{c} - \bm{\xi}) - (\bm{\eta}^T \bm{c}) \bm{\eta}$.

\subsection{Saddle point equations}
\label{apx:spe}

RSA reduce an $(m^2 + 3 m)$-dimensional minimization problem into a $6$-dimensional minimization problem. These six saddle point equations are given by setting derivative $\frac{\partial S}{\partial q_0} = \frac{\partial S}{\partial q_1} = \frac{\partial S}{\partial \hat{q}_0} = \frac{\partial S}{\partial \hat{q}_1} = \frac{\partial S}{\partial r} = \frac{\partial S}{\partial \hat{r}} = 0$ be zero. Among those six saddle point equations, four equations are the same as the saddle point equations in classical generalization error problem \cite{canatar_statistical_2022}:
\begin{align}
    \frac{\partial S}{\partial q_1} = 0, & \Rightarrow \hat{q}_1 = \frac{\beta^2 (q_1 + r^2)}{(\lambda + \beta (q_0 - q_1))^2},  \label{eq:SD-1}\\ 
    \frac{\partial S}{\partial q_0} = 0, & \Rightarrow \hat{q}_0 = \frac{\hat{q}_1}{2} - \frac{1}{2} \frac{\beta}{\lambda + \beta (q_0 - q_1)}, \label{eq:SD-2}\\
    \frac{\partial S}{\partial \hat{q}_0} = 0, & \Rightarrow q_0 =  q_1 + \frac{1}{\beta} \mathrm{Tr} (\mathbf{C} \bm{\Delta}^{- 1}),  \label{eq:SD-4}\\
    \frac{\partial S}{\partial r} = 0, & \Rightarrow \hspace{0.4em} \hat{r} = - \frac{\beta r}{\lambda + \beta (q_0 - q_1)} .  \label{eq:SD-5}
\end{align}
The remaining two saddle points equation $\frac{\partial S}{\partial \hat{q}_1} = \frac{\partial S}{\partial \hat{r}} = 0$ are however quite different. They contain the \textit{contributions from quantum sampling noise} $\mathbf{V}$:
\begin{align}
    \frac{\partial S}{\partial \hat{q}_1} = 0, \Rightarrow~& q_1 = \frac{N \hat{q}_1}{\beta^2} \mathrm{Tr} (\mathbf{C} \bm{\Delta}^{- 1} \mathbf{C} \bm{\Delta}^{- 1}) \nonumber\\
    & \hspace{1.8em} + \left( \bm{W}^{\star} - \frac{N \hat{r}}{\beta} \bm{d} - \frac{N (2 \hat{q}_0 - \hat{q}_1)}{\beta S} \mathbf{V} \bm{c} \right)^T \bm{\Delta}^{- 1} \mathbf{C} \bm{\Delta}^{- 1} \left( \bm{W}^{\star} - \frac{N \hat{r}}{\beta} \bm{d} - \frac{N (2 \hat{q}_0 - \hat{q}_1)}{\beta S} \mathbf{V} \bm{c} \right) \nonumber\\
    & \hspace{1.8em} - \frac{2}{S} \bm{c}^{T} \mathbf{V} \bm{\Delta}^{- 1} \left( \bm{W}^{\star} - \frac{N \hat{r}}{\beta} \bm{d} - \frac{N (2 \hat{q}_0 - \hat{q}_1)}{\beta S} \mathbf{V} \bm{c} \right) + \tilde{\sigma}^2,  \label{eq:SD-3}\\
    \frac{\partial S}{\partial \hat{r}} = 0, \Rightarrow~& r = - \bm{d}^T \bm{\Delta}^{- 1} \left( \bm{W}^{\star} - \frac{N \hat{r}}{\beta} \bm{d} - \frac{N (2 \hat{q}_0 - \hat{q}_1)}{\beta S} \mathbf{V} \bm{c} \right) .  \label{eq:SD-6}
\end{align}
Two commonly appearing terms introduce $\kappa$, the \textit{Signal Capture Threshold} (SCT) 
\begin{align}
    \kappa :=~& \lambda + \beta (q_0 - q_1) = \lambda + \mathrm{Tr} (\mathbf{C} \bm{\Delta}^{- 1}), \\ 
    \frac{2 \hat{q}_0 - \hat{q}_1}{\beta} =~& - \frac{1}{\lambda + \beta (q_0 - q_1)} = - \frac{1}{\kappa} . 
\end{align}
Plugging into $\bm{\Delta} = (\mathbf{I}- \bm{\eta} \bm{\eta}^T) - \frac{N (2 \hat{q}_0 - \hat{q}_1)}{\beta} \mathbf{C}$ yields $\bm{\Delta} = (\mathbf{I}- \bm{\eta} \bm{\eta}^T) + \frac{N}{\kappa} \mathbf{C}$. Thus, we get implicit equation of $\kappa$. 
\begin{equation}
    \kappa = \lambda + \mathrm{Tr} (\mathbf{C} \bm{\Delta}^{- 1}) = \lambda + \mathrm{Tr} \left( \mathbf{C} \bm{} \left( (\mathbf{I}- \bm{\eta} \bm{\eta}^T) + \frac{N}{\kappa} \mathbf{C} \right)^{- 1} \right) .
\end{equation}
The value of $r$ can be solved by Eq.(\ref{eq:SD-5}) and Eq.\,(\ref{eq:SD-6}), (notice $\frac{\hat{r}}{\beta} = - \frac{r}{\lambda + \beta (q_0 - q_1)} = - \frac{r}{\kappa} = \frac{r (2 \hat{q}_0 - \hat{q}_1)}{\beta}$)
\begin{equation}
    r = - \bm{d}^T \bm{\Delta}^{- 1} \left( \bm{W}^{\star} + \frac{N r}{\kappa} \bm{d} + \frac{N}{\kappa S} \mathbf{V} \bm{c} \right) \Rightarrow r = - \frac{\bm{d}^T \bm{\Delta}^{- 1} \left( \bm{W}^{\star} + \frac{N}{\kappa S} \mathbf{V} \bm{c} \right)}{1 + \frac{N}{\kappa} \bm{d}^T \bm{\Delta}^{- 1} \bm{d}}, \quad \mathrm{and} \quad \hat{r} = - \frac{\beta r}{\kappa} .
\end{equation}
By looking at the expression of $r$, we define for convenience that
\begin{equation}
    \widetilde{\bm{W}}^{\star} = \bm{W}^{\star} + \frac{N}{\kappa S} \mathbf{V} \bm{c} . \label{eq:tildeWstar}
\end{equation}
Solving all remaining four saddle point equations yields
\begin{eqnarray}
    & q_1 = \frac{\frac{N r^2}{\kappa^2} \mathrm{Tr} (\mathbf{C} \bm{\Delta}^{- 1} \mathbf{C} \bm{\Delta}^{- 1}) + \left( \widetilde{\bm{W}}^{\star} + \frac{N r}{\kappa} \bm{d} \right)^T \bm{\Delta}^{- 1} \mathbf{C} \bm{\Delta}^{- 1} \left( \widetilde{\bm{W}}^{\star} + \frac{N r}{\kappa} \bm{d} \right) - \frac{2}{S} \bm{c}^{T} \mathbf{V} \bm{\Delta}^{- 1} \left( \widetilde{\bm{W}}^{\star} + \frac{N r}{\kappa} \bm{d} \right)}{1 - \frac{N}{\kappa^2} \mathrm{Tr} (\mathbf{C} \bm{\Delta}^{- 1} \mathbf{C} \bm{\Delta}^{- 1})}, & \nonumber\\
    & q_0 = q_1 + \frac{\kappa - \lambda}{\beta}, \quad \hat{q}_1 = \frac{\beta^2 (q_1 + r^2)}{\kappa^2}, \quad \hat{q}_0 = \frac{1}{2} \left( \hat{q}_1 - \frac{\beta}{\kappa} \right). & 
\end{eqnarray}
Finally, by combining solutions to $q_0, q_1, \hat{q}_0, \hat{q}_1, r, \hat{r}$, and using the \textit{Sherman--Morrison formula} $\left( \bm{\Delta} + \frac{N}{\kappa} \bm{d}^T \bm{d} \right)^{- 1} = \bm{\Delta}^{- 1} - \frac{N}{\kappa} \frac{\bm{\Delta}^{- 1} \bm{d} \bm{d}^T \bm{\Delta}^{- 1}}{1 + \frac{N}{\kappa} \bm{d}^T \bm{\Delta}^{- 1} \bm{d}}$, the replica-symmetry free energy is
\begin{equation}
    \mathbb{E}_{\mathcal{D}} [\ln Z [\bm{\xi}, \bm{\eta}, \beta ]] = \frac{N}{2} \left[ \frac{\kappa - \lambda}{\kappa} - \ln \left( \frac{\kappa}{\lambda} \right) - \frac{\beta \tilde{\sigma}^2}{\kappa} \right] - \frac{1}{2} \ln \det (\bm{\Delta}) + \frac{\beta}{2} [\widetilde{\bm{W}}^{\star T} \widetilde{\bm{\Delta}}^{- 1} \widetilde{\bm{W}}^{\star} - (\bm{W}^{\star} - \bm{\xi})^T \bm{c}] \label{eq:RSFE}
\end{equation}
where we recall $\bm{\Delta} =\mathbf{I}+ \frac{N}{\kappa} \mathbf{C}- \bm{\eta} \bm{\eta}^T$, $\mathbf{C}=\mathbf{G}+ \frac{1}{S} \mathbf{V}- \bm{d} \bm{d}^T$, $\bm{W}^{\star} = (\bm{c} - \bm{\xi}) - (\bm{\eta}^T \bm{c}) \bm{\eta}$, $\widetilde{\bm{W}}^{\star} = \bm{W}^{\star} + \frac{N}{\kappa S} \mathbf{V} \bm{c}$, and finally $\kappa$ is solved self-consistently through $\kappa = \lambda + \mathrm{Tr} (\mathbf{C} \bm{\Delta}^{- 1})$. We also newly define the matrix
\begin{equation}
    \widetilde{\bm{\Delta}} = \bm{\Delta} + \frac{N}{\kappa} \bm{d}^T \bm{d} =\mathbf{I}+ \frac{N}{\kappa} \left( \mathbf{G}+ \frac{1}{S} \mathbf{V} \right) - \bm{\eta} \bm{\eta}^T, \label{eq:tildeDelta}
\end{equation}

\noindent \textbf{Remark}. The correctness of Eq.\,(\ref{eq:RSFE}) can be checked by taking $S \rightarrow \infty$. Under this limit, Eq.\,(\ref{eq:tildeWstar}) becomes $\widetilde{\bm{W}}^{\star} = \bm{W}^{\star} + \frac{N}{\kappa S} \mathbf{V} \bm{c} \rightarrow \bm{W}^{\star}$ and Eq.\,(\ref{eq:tildeDelta}) becomes $\widetilde{\bm{\Delta}} =\mathbf{I}+ \frac{N}{\kappa} \left( \mathbf{G}+ \frac{1}{S} \mathbf{V} \right) - \bm{\eta} \bm{\eta}^T \rightarrow \mathbf{I}+ \frac{N}{\kappa} \mathbf{G}- \bm{\eta} \bm{\eta}^T$, then it is direct to see that Eq.\,(\ref{eq:RSFE}) reduces to Eq.\,(F67) in Ref.\,\cite{canatar_statistical_2022}.

\section{Training and Generalization Error under Original Features}
\label{apx:Original_Features}

We recall that Eq.\,(\ref{eq:EDwast-1-old}, \ref{eq:EDwast-2-old}) indicates the first and second order moments of trained weights can be expressed in terms of the derivative with respect the added source in partition function $Z [\bm{\xi}, \bm{\eta}, \beta ]$ at $\bm{\xi}, \bm{\eta} = \bm{0}$:
\begin{align}
    \mathbb{E}_{\mathcal{D}} [w_k^{\ast}] & = \lim_{\beta \rightarrow \infty} \frac{1}{\beta} \frac{\partial}{\partial \xi_k} \mathbb{E}_{\mathcal{D}} [\ln Z [\bm{\xi}, \bm{\eta}, \beta ]] |_{\bm{\xi}, \bm{\eta} = \bm{0}}, \label{eq:EDwast-1-new} \\
    \mathbb{E}_{\mathcal{D}} [w_j^{\ast} w_k^{\ast}] & = \lim_{\beta \rightarrow \infty} \left. \frac{1}{\beta} \frac{\partial^2}{\partial \eta_j \partial \eta_k} \mathbb{E}_{\mathcal{D}} [\ln Z [\bm{\xi}, \bm{\eta}, \beta ]] \right|_{\bm{\xi}, \bm{\eta} = \bm{0}} . \label{eq:EDwast-2-new}
\end{align}

\subsection{First order moments of trained weights}

Computation of first order moments of trained weights is relatively easier. At $\bm{\xi}, \bm{\eta} = \bm{0}$, we have
\begin{equation}
   \bm{\Delta} |_{\bm{\xi}, \bm{\eta} = \bm{0}} =\mathbf{I}+ \frac{N}{\kappa} \left( \mathbf{G}+ \frac{1}{S} \mathbf{V}- \bm{d} \bm{d}^T \right), \quad  \widetilde{\bm{\Delta}} |_{\bm{\xi}, \bm{\eta} = \bm{0}} =\mathbf{I}+ \frac{N}{\kappa} \left( \mathbf{G}+ \frac{1}{S} \mathbf{V} \right) .
\end{equation}
The first order moment is computed from Eq.\,(\ref{eq:RSFE}) and Eq.\,(\ref{eq:EDwast-1-new}) (we may abbreviate $ \widetilde{\bm{\Delta}}^{- 1} = \widetilde{\bm{\Delta}}^{- 1} |_{\bm{\xi}, \bm{\eta} = \bm{0}}$ when no confusion):
\begin{align}
    \mathbb{E}_{\mathcal{D}} [\bm{w}^{\ast} ] = \left( \mathbf{I}- \widetilde{\bm{\Delta}}^{- 1} \left( \mathbf{I}+  \frac{N}{\kappa S} \mathbf{V} \right) \right) \bm{c} = \left( \frac{\kappa}{N} \mathbf{I}+ \left( \mathbf{G}+ \frac{\mathbf{V}}{S}  \right) \right)^{- 1} \mathbf{G} \bm{c} .
\end{align}
where the self-consistent equation of $\kappa$ is 
\begin{equation}
    \kappa = \lambda + \kappa \sum_{k > 0} \frac{1 + \frac{\beta_k^2}{S}}{N \left( 1 + \frac{\beta_k^2}{S} \right) + \kappa} .
\end{equation}
This is a reasonable result, because when taking $N \rightarrow \infty$, we get:
\begin{equation}
    \lim_{N \rightarrow \infty} \mathbb{E}_{\mathcal{D}} [\bm{w}^{\ast} ] = \left( \mathbf{G}+ \frac{\mathbf{V}}{S} \right)^{- 1} \mathbf{G} \bm{c} .
\end{equation}
This is exactly the biased trained weights by minimizing the quadratic loss for large dataset size limit $N \to \infty$ \cite{hu_tackling_2023}.

\subsection{Second order moments of trained weights}

The first order derivatives $  \partial_{\bm{\eta}} \bm{W}^{\star} |_{\bm{} \bm{\eta} = \bm{0}} = \partial_{\bm{\eta}} \kappa |_{\bm{} \bm{\eta} = \bm{0}} =  \partial_{\bm{\eta}} \bm{\Delta} |_{\bm{} \bm{\eta} = \bm{0}} =  \partial_{\bm{\eta}} \mu |_{\bm{} \bm{\eta} = \bm{0}} = 0$ vanishes. Now we can abbreviate $\partial_k = \frac{\partial}{\partial \eta_k}$. The second order derivatives of $\kappa$ and $\widetilde{\bm{\Delta}}$ gives
\begin{align}
   \partial_j \partial_k \kappa |_{\bm{\xi}, \bm{\eta} = \bm{0}} & = \frac{2}{1 - \gamma} (\bm{\Delta}^{- 1} \mathbf{C} \bm{\Delta}^{- 1})_{j k}, \\ 
   \partial_j \partial_k \widetilde{\bm{\Delta}} |_{\bm{\xi}, \bm{\eta} = \bm{0}} & = - \frac{N}{\kappa^2} \left( \mathbf{G}+ \frac{1}{S} \mathbf{V} \right) \partial_j \partial_k \kappa - \partial_j \partial_k (\bm{\eta} \bm{\eta}^T) . 
\end{align}
where we define
\begin{equation}
    \gamma = \frac{N}{\kappa^2} \mathrm{Tr} (\mathbf{C} \bm{\Delta}^{- 1} \mathbf{C} \bm{\Delta}^{- 1}) .
\end{equation}
Then the second order derivatives of $\widetilde{\bm{W}}^{\star T} \widetilde{\bm{\Delta}}^{- 1} \widetilde{\bm{W}}^{\star}$ and $- \frac{N \tilde{\sigma}^2}{\kappa} - (\bm{W}^{\star} - \bm{\xi})^T \bm{c}$
\begin{align}
    \left. \frac{\partial^2}{\partial \eta_j \partial \eta_k} (\widetilde{\bm{W}}^{\star T} \widetilde{\bm{\Delta}}^{- 1} \widetilde{\bm{W}}^{\star}) \right|_{\bm{\xi},
    \bm{\eta} = \bm{0}} =~& - 2 \bm{c}^{T} \left( \mathbf{I}+ \frac{N}{\kappa S} \mathbf{V} \right) \widetilde{\bm{\Delta}}^{- 1} \left( \partial_j \partial_k (\bm{\eta} \bm{\eta}^T) + \frac{N}{\kappa^2 S} \mathbf{V} \partial_j \partial_k \kappa \right) \bm{c} \nonumber\\
    & + \frac{N}{\kappa^2} \partial_j \partial_k \kappa \bm{c}^{T} \left( \mathbf{I}+ \frac{N}{\kappa S} \mathbf{V} \right) \widetilde{\bm{\Delta}}^{- 1} \left( \mathbf{G}+ \frac{1}{S} \mathbf{V} \right) \widetilde{\bm{\Delta}}^{- 1} \left( \mathbf{I}+ \frac{N}{\kappa S} \mathbf{V} \right) \bm{c} \nonumber\\
    & + \bm{c}^{T} \left( \mathbf{I}+ \frac{N}{\kappa S} \mathbf{V} \right) \widetilde{\bm{\Delta}}^{- 1} \partial_j \partial_k (\bm{\eta} \bm{\eta}^T) \widetilde{\bm{\Delta}}^{- 1} \left( \mathbf{I}+ \frac{N}{\kappa S} \mathbf{V} \right) \bm{c}, \\
    \left. \frac{\partial^2}{\partial \eta_j \partial \eta_k} \left( - \frac{N \tilde{\sigma}^2}{\kappa} - (\bm{W}^{\star} - \bm{\xi})^T \bm{c} \right) \right|_{\bm{\xi}, \bm{\eta} =
    \bm{0}} =~& \frac{N \tilde{\sigma}^2}{\kappa^2} \partial_j \partial_k \kappa + \bm{c}^{T} \partial_j \partial_k (\bm{\eta} \bm{\eta}^T) \bm{c} . 
\end{align}
We now plug Eq.\,(\ref{eq:EDwast-2-new}) to Eq.\,(\ref{eq:RSFE}), and collect terms involving $\partial_j \partial_k \kappa$ and $\partial_j \partial_k (\bm{\eta} \bm{\eta}^T)$ separately. It gives the second order moment
\begin{align}
    \mathbb{E}_{\mathcal{D}} [\bm{w}^{\ast} \bm{w}^{\ast T}] =~& \frac{N (\bm{\Delta}^{- 1} \mathbf{C} \bm{\Delta}^{- 1})}{\kappa^2 (1 - \gamma)} \left( \tilde{\sigma}^2 + \bm{c}^{T} \left( \mathbf{I}+ \frac{N}{\kappa S} \mathbf{V} \right) \widetilde{\bm{\Delta}}^{- 1} \left( \mathbf{G}+ \frac{1}{S} \mathbf{V} \right) \widetilde{\bm{\Delta}}^{- 1} \left( \mathbf{I}+ \frac{N}{\kappa S} \mathbf{V} \right) \bm{c} \right) \nonumber\\
    & - \frac{N (\bm{\Delta}^{- 1} \mathbf{C} \bm{\Delta}^{- 1})}{\kappa^2 (1 - \gamma)} \left( 2 \bm{c}^{T} \left( \mathbf{I}+ \frac{N}{\kappa S} \mathbf{V} \right) \widetilde{\bm{\Delta}}^{- 1} \frac{\mathbf{V}}{S} \bm{c} \right) +\mathbb{E}_{\mathcal{D}} [\bm{w}^{\ast}] \mathbb{E}_{\mathcal{D}} [\bm{w}^{\ast T}].
\end{align}

\subsection{Training and generalization errors}

We first notice that
\begin{equation}
    - 2 \lim_{\beta \rightarrow \infty} \frac{\partial}{\partial \beta} \mathbb{E}_{\mathcal{D}} [\ln Z [\bm{\xi}, \bm{\eta}, \beta ]] |_{\bm{\xi}, \bm{\eta} = \bm{0}} = \frac{N \tilde{\sigma}^2}{\kappa} - \bm{c}^{T} \left( \mathbf{I}+ \frac{N}{\kappa S} \mathbf{V} \right) \widetilde{\bm{\Delta}}^{- 1} \left( \mathbf{I}+ \frac{N}{\kappa S} \mathbf{V} \right) \bm{c} + \bm{c}^{T} \bm{c} .
\end{equation}
For simplicity, we abbreviate $\delta = \frac{1}{\kappa^2} \mathrm{Tr} (\bm{\Delta}^{- 1} \mathbf{C} \bm{\Delta}^{- 1})$. Then the $\mathcal{D}$-average of optimized weights norm-sqaure is
\begin{align}
    \mathbb{E}_{\mathcal{D}} [\| \bm{w}^{\ast} \|^2] =~& \frac{N \delta}{1 - \gamma} \tilde{\sigma}^2 + \left( \frac{\delta \kappa}{1 - \gamma} - 2 \right) \bm{c}^{T} \left( \mathbf{I}+ \frac{N}{\kappa S} \mathbf{V} \right) \widetilde{\bm{\Delta}}^{- 1} \left( \mathbf{I}+ \frac{N}{\kappa S} \mathbf{V} \right) \bm{c} \nonumber\\
    & - \frac{2 N}{\kappa} \left( \frac{\delta \kappa}{1 - \gamma} - 1 \right) \left( 2 \bm{c}^{T} \left( \mathbf{I}+ \frac{N}{\kappa S} \mathbf{V} \right) \widetilde{\bm{\Delta}}^{- 1} \frac{\mathbf{V}}{S} \bm{c} \right) \nonumber\\
    & + \left( - \frac{\delta \kappa}{1 - \gamma} + 1 \right) \bm{c}^{T} \left( \mathbf{I}+ \frac{N}{\kappa S} \mathbf{V} \right) \widetilde{\bm{\Delta}}^{- 2} \left( \mathbf{I}+ \frac{N}{\kappa S} \mathbf{V} \right) \bm{c} + \bm{c}^{T} \bm{c} . 
\end{align}
By using the relation $\delta \kappa + (\gamma - 1) = - \frac{\lambda}{\kappa}$ \cite{canatar_statistical_2022}, we obtain the explicit expression of average training error
\begin{align}
    E_t & = \frac{2 \lambda}{N} \left( - \lim_{\beta \rightarrow \infty} \left. \frac{\partial}{\partial \beta} \langle \ln Z [\bm{\xi}, \bm{\eta}, \beta ] \rangle_{\mathcal{D}} \right|_{\bm{\xi}, \bm{\eta} = \bm{0}} - \frac{1}{2} \langle \| \bm{w}^{\ast} \|^2 \rangle_{\mathcal{D}} \right) \nonumber\\
    & = \frac{\lambda^2}{\kappa^2 (1 - \gamma)} \left( \tilde{\sigma}^2 + \bm{c}^{T} \left( \mathbf{I}+ \frac{N}{\kappa S} \mathbf{V} \right) \widetilde{\bm{\Delta}}^{- 1} \left( \mathbf{G}+ \frac{1}{S} \mathbf{V} \right) \widetilde{\bm{\Delta}}^{- 1} \left( \mathbf{I}+ \frac{N}{\kappa S} \mathbf{V} \right) \bm{c} - 2 \bm{c}^{T} \left( \mathbf{I}+ \frac{N}{\kappa S} \mathbf{V} \right) \widetilde{\bm{\Delta}}^{- 1} \frac{\mathbf{V}}{S} \bm{c} \right) . 
\end{align}
Eventually, computation of average generalization error is more direct:
\begin{align}
    E_g =~& \mathrm{Tr} \left( \left( \mathbf{G}+ \frac{1}{S} \mathbf{V} \right) \langle \bm{w}^{\ast} \bm{w}^{\ast T} \rangle_{\mathcal{D}} \right) - 2 \bm{c}^{T} \mathbf{G} \langle \bm{w}^{\ast} \rangle_{\mathcal{D}} + 1 \nonumber\\
    =~& \mathrm{Tr} \left( \left( \mathbf{G}+ \frac{1}{S} \mathbf{V} \right) \left( \frac{N (\bm{\Delta}^{- 1} \mathbf{C} \bm{\Delta}^{- 1})}{\kappa^2 (1 - \gamma)} \left( \tilde{\sigma}^2 + \bm{c}^{T} \left( \mathbf{I}+ \frac{N}{\kappa S} \mathbf{V} \right) \widetilde{\bm{\Delta}}^{- 1} \left( \mathbf{G}+ \frac{1}{S} \mathbf{V} \right) \widetilde{\bm{\Delta}}^{- 1} \left( \mathbf{I}+ \frac{N}{\kappa S} \mathbf{V} \right) \bm{c} \right) \right) \right) \nonumber\\
    & - \mathrm{Tr} \left( \left( \mathbf{G}+ \frac{1}{S} \mathbf{V} \right) \left( \frac{N (\bm{\Delta}^{- 1} \mathbf{C} \bm{\Delta}^{- 1})}{\kappa^2 (1 - \gamma)} \left( 2 \bm{c}^{T} \left( \mathbf{I}+ \frac{N}{\kappa S} \mathbf{V} \right) \widetilde{\bm{\Delta}}^{- 1} \frac{\mathbf{V}}{S} \bm{c} \right) \right) \right) \nonumber\\
    & + \mathrm{Tr} \left( \left( \mathbf{G}+ \frac{1}{S} \mathbf{V} \right) \left( \frac{\kappa}{N} \mathbf{I}+ \left( \mathbf{G}+ \frac{\mathbf{V}}{S} \right) \right)^{- 1} \mathbf{G} \bm{c} \bm{c}^{T} \mathbf{G} \left( \frac{\kappa}{N} \mathbf{I}+ \left( \mathbf{G}+ \frac{\mathbf{V}}{S} \right) \right)^{- 1} \right) \nonumber\\
    & - 2 \bm{c}^{T} \mathbf{G} \left( \frac{\kappa}{N} \mathbf{I}+ \left( \mathbf{G}+ \frac{\mathbf{V}}{S} \right) \right)^{- 1} \mathbf{G} \bm{c} + 1, 
\end{align}
where $\tilde{\sigma}^2 = \frac{1}{S} \bm{c}^{T} \mathbf{V} \bm{c} +\mathbf{E}_{\bm{u}} [f^2_{\bot}] = \bm{c}^{T} \left( -\mathbf{G}+ \frac{1}{S} \mathbf{V} \right) \bm{c} + 1$.

\section{Transform to Eigentask Basis}
\label{apx:Eigentask}

Recall in the main text, we decompose the target function $f^\star$ under basis of both original features Eq.\,(\ref{eq:cx}) and eigentasks Eq.\,(\ref{eq:ay}).
\begin{align}
    f^{\star} = \bm{c} \cdot \bm{x} + f_{\bot} = \bm{a} \cdot \bm{y} + f_{\bot}, 
\end{align}
Also, under eigentask basis, we have
\begin{align}
    \mathbf{G} = \mathbf{I}, \quad \mathbf{V} = \mathrm{diag} (\{ \beta_k^2 \}), \quad \bm{d} = (1, 0, \cdots, 0)^T . 
\end{align}
Then the average training and generalization errors are
\begin{align}
    E_g & = \frac{\gamma}{1 - \gamma} \left( \mathbb{E}_{\bm{u}} [f^2_{\bot}] + \sum_k a_k^{2} \left( \frac{\left( \frac{\beta_k^2}{S} + \frac{\kappa}{N} \right)^2 + \frac{\beta_k^2}{S}}{\left( 1 + \frac{\beta_k^2}{S} + \frac{\kappa}{N} \right)^2} \right) \right) + \sum_k a_k^{2} \left( \frac{1+\frac{\beta^2_k}{S}}{\left( 1 + \frac{\beta_k^2}{S} + \frac{\kappa}{N} \right)^2} - \frac{2}{1 + \frac{\beta_k^2}{S} + \frac{\kappa}{N}} \right) + 1 \nonumber\\
    & = \frac{1}{1 - \gamma} \left( \mathbb{E}_{\bm{u}} [f^2_{\bot}] + \sum_k a_k^{2} \left( \frac{\left( \frac{\beta_k^2}{S} + \frac{\kappa}{N} \right)^2 + \frac{\beta_k^2}{S}}{\left( 1 + \frac{\beta_k^2}{S} + \frac{\kappa}{N} \right)^2} \right) \right), \label{eq:Egapx} \\
    E_t & = \frac{\lambda^2}{\kappa^2 (1 - \gamma)} \left( \mathbb{E}_{\bm{u}} [f^2_{\bot}] + \sum_k a_k^{2} \left( \frac{\left( \frac{\beta_k^2}{S} + \frac{\kappa}{N} \right)^2 + \frac{\beta_k^2}{S}}{\left( 1 + \frac{\beta_k^2}{S} + \frac{\kappa}{N} \right)^2} \right) \right) = \frac{\lambda^2}{\kappa^2} E_g . \label{eq:Etapx} 
\end{align}
where
\begin{align}
    \kappa & = \lambda + \kappa \sum_{k > 1} \frac{1 + \frac{\beta_k^2}{S}}{N \left( 1 + \frac{\beta_k^2}{S} \right) + \kappa}, \label{eq:SCT}\\
    \gamma & = \sum_{k > 1} \frac{N \left( 1 + \frac{\beta_k^2}{S} \right)^2}{\left( N \left( 1 + \frac{\beta_k^2}{S} \right) + \kappa \right)^2} . \label{eq:gamma}
\end{align}
One important scenario is the small regularization limit $\lambda \rightarrow 0$. For sufficiently large data set $N \rightarrow \infty$, it is easy to verify that both Eq.\,(\ref{eq:Egapx}) and Eq.\,(\ref{eq:Etapx}) converge to
\begin{equation}
    \lim_{N \rightarrow \infty} E_t (\lambda = 0) = \lim_{N \rightarrow \infty} E_g (\lambda = 0) = 1 - \sum_k \frac{a_k^{2}}{1 + \frac{\beta_k^2}{S}}.
\end{equation}
We also notice that Eq.\,(\ref{eq:SCT}) should have completely different behaviour for $N > K - 1$ and $N \leq K - 1$:
\begin{itemize}
    \item If $N > K - 1$: Suppose the equation of $\kappa$ has a positive solution, then
    \begin{equation}
        1 = \sum_{k > 0} \frac{1 + \frac{\beta_k^2}{S}}{N \left( 1 + \frac{\beta_k^2}{S} \right) + \kappa} < \sum_{k > 0} \frac{1 + \frac{\beta_k^2}{S}}{N \left( 1 + \frac{\beta_k^2}{S} \right)} = \frac{K - 1}{N} < 1,
    \end{equation}
    which results in a contradiction. It means that $\kappa = 0$ for $N < K - 1$. Similarly,
    \begin{equation}
        \gamma = \sum_{k > 0} \frac{N \left( 1 + \frac{\beta_k^2}{S} \right)^2}{\left( N \left( 1 + \frac{\beta_k^2}{S} \right) + \kappa \right)^2} = \sum_{k > 0} \frac{N \left( 1 + \frac{\beta_k^2}{S} \right)^2}{\left( N \left( 1 + \frac{\beta_k^2}{S} \right) \right)^2} = \frac{K - 1}{N} .
    \end{equation}
    \item If $N \leq K - 1$: Now $\kappa$ has non-zero solution $\kappa = \kappa (N) > 0$, and $\gamma = \gamma (N) < \frac{K - 1}{N}$.
\end{itemize}
The non-analytical behaviour explains the phase transition phenomenon in Fig.\,\ref{fig:1} of the main text. 

\stopcontents[appendices]
\end{widetext}

\end{document}